%%%%%%%%%%%%%%% FORMATO
\magnification=\magstep1
\hoffset=0.5truecm
\voffset=0.5truecm
\hsize=16.5truecm 
\vsize=22.0truecm
\baselineskip=14pt plus0.1pt minus0.1pt 
\parindent=25pt
\lineskip=4pt\lineskiplimit=0.1pt      
\parskip=0.1pt plus1pt

\let\st=\scriptstyle

%%%%%%%%%%%%%%%%%%%%%%%%%%%  FONTS

\font\ninerm=cmr9

\font\ninebf=cmbx9

\font\sixrm=cmr6

%%%%%%%%%%%%%%%%%%%%%%%%% GRECO

\let\a=\alpha \let\b=\beta   \let\d=\delta  
 \let\g=\gamma \let\h=\eta      \let\l=\lambda
\let\m=\mu   \let\n=\nu         
\let\r=\rho  \let\s=\sigma \let\t=\tau   
  
\let\D=\Delta     \let\L=\Lambda 
\let\O=\Omega      
\let\Y=\Upsilon

%%%%%%%%%%%%%%%%%%%%%%% CALLIGRAFICHE
%
  \def\cC{{\cal C}} \def\cD{{\cal D}}
\def\cE{{\cal E}}   
\def\cI{{\cal I}}  \def\cK{{\cal K}} \def\cL{{\cal L}}
   
  \def\cS{{\cal S}} \def\cT{{\cal T}}
\def\cU{{\cal U}}

%%%%%%%%%%%%%%%%%%%%%%%%%%%%%%%%%% figure
%
\newdimen\xshift 
\newdimen\yshift
\newdimen\xwidth 
\def\eqfig#1#2#3#4#5#6{
  \par\xwidth=#1 \xshift=\hsize \advance\xshift 
  by-\xwidth \divide\xshift by 2
  \yshift=#2 \divide\yshift by 2
  \vbox{
  \line{\hglue\xshift \vbox to #2{
    \smallskip
    \vfil#3 
    \includegraphics{#4.ps}
    }
    \hfill\raise\yshift\hbox{#5}
  }
  \smallskip
  \centerline{#6}
  }
  \smallskip
}
%

%%%%%%%%%%%%%%%%%%%%%  Numerazione pagine
%
\def\data{\number\day/\ifcase\month\or gennaio \or febbraio \or marzo \or
aprile \or maggio \or giugno \or luglio \or agosto \or settembre
\or ottobre \or novembre \or dicembre \fi/\number\year}

%%\newcount\tempo
%%\tempo=\number\time\divide\tempo by 60}

\setbox200\hbox{$\scriptscriptstyle \data $}

\newcount\pgn 
\pgn=1
\def\foglio{\veroparagrafo:\number\pgn
\global\advance\pgn by 1}

%%%%%%%%%%%%%%%%% EQUAZIONI E TEOREMI CON NOMI SIMBOLICI

\global\newcount\numsec
\global\newcount\numfor
\global\newcount\numfig
\global\newcount\numtheo

\gdef\profonditastruttura{\dp\strutbox}

\def\senondefinito#1{\expandafter\ifx\csname#1\endcsname\relax}

\def\SIA #1,#2,#3 {\senondefinito{#1#2}%
   \expandafter\xdef\csname #1#2\endcsname{#3}\else
   \write16{???? ma #1,#2 e' gia' stato definito !!!!}\fi}

\def\etichetta(#1){(\veroparagrafo.\veraformula)
   \SIA e,#1,(\veroparagrafo.\veraformula)
   \global\advance\numfor by 1
   \write15{\string\FU (#1){\equ(#1)}}
   \write16{ EQ \equ(#1) == #1  }}

\def\FU(#1)#2{\SIA fu,#1,#2 }

\def\tetichetta(#1){{\veroparagrafo.\verotheo}%
   \SIA theo,#1,{\veroparagrafo.\verotheo}
   \global\advance\numtheo by 1%
   \write15{\string\FUth (#1){\thm[#1]}}%
   \write16{ TH \thm[#1] == #1  }}

\def\FUth(#1)#2{\SIA futh,#1,#2 }

\def\getichetta(#1){Fig. \verafigura
 \SIA e,#1,{\verafigura}
 \global\advance\numfig by 1
 \write15{\string\FU (#1){\equ(#1)}}
 \write16{ Fig. \equ(#1) ha simbolo  #1  }}

\newdimen\gwidth

\def\BOZZA{
 \def\alato(##1){
 {\vtop to \profonditastruttura{\baselineskip
 \profonditastruttura\vss
 \rlap{\kern-\hsize\kern-1.2truecm{$\scriptstyle##1$}}}}}
 \def\galato(##1){ \gwidth=\hsize \divide\gwidth by 2
 {\vtop to \profonditastruttura{\baselineskip
 \profonditastruttura\vss
 \rlap{\kern-\gwidth\kern-1.2truecm{$\scriptstyle##1$}}}}}
 \def\talato(##1){\rlap{\sixrm\kern -1.2truecm ##1}}
}

\def\alato(#1){}
\def\galato(#1){}
\def\talato(#1){}

\def\veroparagrafo{\ifnum\numsec<0 A\number-\numsec\else
   \number\numsec\fi}
\def\veraformula{\number\numfor}
\def\verotheo{\number\numtheo}
\def\verafigura{\number\numfig}

\def\Thm[#1]{\tetichetta(#1)}
\def\thf[#1]{\senondefinito{futh#1}$\clubsuit$[#1]\else
   \csname futh#1\endcsname\fi}
\def\thm[#1]{\senondefinito{theo#1}$\spadesuit$[#1]\else
   \csname theo#1\endcsname\fi}

\def\Eq(#1){\eqno{\etichetta(#1)\alato(#1)}}
\def\eq(#1){\etichetta(#1)\alato(#1)}
\def\eqv(#1){\senondefinito{fu#1}$\clubsuit$(#1)\else
   \csname fu#1\endcsname\fi}
\def\equ(#1){\senondefinito{e#1}$\spadesuit$(#1)\else
   \csname e#1\endcsname\fi}
\let\eqf=\eqv

% -------------------------------------------------------------------------
%
%  Numerazione verso il futuro ed eventuali paragrafi
%  precedenti non inseriti nel file da compilare
%
\def\include#1{
\openin13=#1.aux \ifeof13 \relax \else
\input #1.aux \closein13 \fi}
\openin14=\jobname.aux \ifeof14 \relax \else
\input \jobname.aux \closein14 \fi
\openout15=\jobname.aux

% -------------------------------------------------------------------------
%
% 

% -------------------------------------------------------------------------
%
\footline={\rlap{\hbox{\copy200}\ $\st[\number\pageno]$}\hss\tenrm
\foglio\hss}

% ---------------- fonti disponibili ---------------------------
%
\newcount\fnts
\fnts=0
\fnts=1 %-----comment if fonts msam, msbm, eufm are not available

%------------------------- Altre macro da chiamare ------------
%

\def\smallno{\smallskip\noindent}
\def\medno{\medskip\noindent}
\def\bigno{\bigskip\noindent}

\def\thsp{\thinspace}

\def\tthsp{\kern .083333 em}

\def\?{\mskip -10mu}

%------------------------ itemizing
%

\def\indbox#1{\hbox to \parindent{\hfil\ #1\hfil} }

\def\ref[#1]{[#1]}

\def\beginsubsection#1\par{\bigskip\leftline{\it #1}\nobreak\smallskip
	    \noindent}

\newfam\msafam
\newfam\msbfam
\newfam\eufmfam

% -------------------------------------------------- math macros --------
%
\ifnum\fnts=0% ------------Se non ci sono le fonti
  \def\bZ{ { {\rm Z} \mskip -6.6mu {\rm Z} }  }
  \def\bN{ { {\rm N} \mskip -6.6mu {\rm N} }  }
  \def\bR{{\rm I\!R}}
  \def\bb{ \vrule height 6.7pt width 0.5pt depth 0pt }
  \def\bC{ { {\rm C} \mskip -8mu \bb \mskip 8mu } }
  \def\bE{{\rm I\!E}}
  \def\bP{{{\rm I\!P}}
  \def\mbox{
  \vbox{ \hrule width 6pt
     \hbox to 6pt{\vrule\vphantom{k} \hfil\vrule}
     \hrule width 6pt}
  }
  \def\QED{\ifhmode\unskip\nobreak\fi\quad
    \ifmmode\mbox\else$\mbox$\fi}
  \let\restriction=\lceil
\else% ------ o se ci sono
  \def\hexnumber#1{%
  \ifcase#1 0\or 1\or 2\or 3\or 4\or 5\or 6\or 7\or 8\or
  9\or A\or B\or C\or D\or E\or F\fi}
  \font\tenmsa=msam10
  \font\sevenmsa=msam7
  \font\fivemsa=msam5
  \textfont\msafam=\tenmsa
  \scriptfont\msafam=\sevenmsa
  \scriptscriptfont\msafam=\fivemsa        
  \edef\msafamhexnumber{\hexnumber\msafam}%
  \mathchardef\restriction"1\msafamhexnumber16
  \mathchardef\square"0\msafamhexnumber03
  \def\QED{\ifhmode\unskip\nobreak\fi\quad
    \ifmmode\square\else$\square$\fi}            
  \font\tenmsb=msbm10
  \font\sevenmsb=msbm7
  \font\fivemsb=msbm5
  \textfont\msbfam=\tenmsb
  \scriptfont\msbfam=\sevenmsb
  \scriptscriptfont\msbfam=\fivemsb
  \def\Bbb#1{\fam\msbfam\relax#1}    
  \font\teneufm=eufm10
  \font\seveneufm=eufm7
  \font\fiveeufm=eufm5
  \textfont\eufmfam=\teneufm
  \scriptfont\eufmfam=\seveneufm
  \scriptscriptfont\eufmfam=\fiveeufm
  \def\frak#1{{\fam\eufmfam\relax#1}}
  \let\goth\frak
  \def\bZ{{\Bbb Z}}
  \def\bN{{\Bbb N}}
  \def\bR{{\Bbb R}}
  \def\bC{{\Bbb C}}
  \def\bE{{\Bbb E}}
  \def\bP{{\Bbb P}}
  \def\bI{{\Bbb I}}
\fi
%
%-------------------------------------------------------------------
%
% ------- Per compatibilita'
%
\let\integer=\bZ
\let\real=\bR
\let\complex=\bC
\let\Ee=\bE
\let\Pp=\bP
\let\Dir=\cE
\let\Z=\integer
\let\uline=\underline
\def\Zp{{\integer_+}}
\def\ZpN{{\integer_+^N}}
\def\ZZ{{\integer^2}}
\def\ZZt{\integer^2_*}
\let\neper=e
\let\ii=i
\let\mmin=\wedge
\let\mmax=\vee
\def\identity{ {1 \mskip -5mu {\rm I}}  }
\def\ie{\hbox{\it i.e.\ }}
\let\id=\identity
\let\emp=\emptyset
\let\sset=\subset
\def\ssset{\subset\subset}
\let\setm=\setminus
\def\nep#1{ \neper^{#1}}
\let\uu=\underline
\def\ov#1{{1\over#1}}
\let\nea=\nearrow
\let\dnar=\downarrow
\let\imp=\Rightarrow
\let\de=\partial
\def\dep{\partial^+}
\def\deb{\bar\partial}
\def\tc{\thsp | \thsp}
\let\<=\langle
\let\>=\rangle
\def\xx{ {\{x\}} }
\def\xy{ { \{x,y\} } }
\def\pmu{\{-1,1\}}
\def\Pro{\noindent{\it Proof.}}
\def\sump{\mathop{{\sum}'}}
\def\tr{ \mathop{\rm tr}\nolimits }
\def\intt{ \mathop{\rm int}\nolimits }
\def\ext{ \mathop{\rm ext}\nolimits }
\def\Tr{ \mathop{\rm Tr}\nolimits }
\def\ad{ \mathop{\rm ad}\nolimits }
\def\Ad{ \mathop{\rm Ad}\nolimits }
\def\dim{ \mathop{\rm dim}\nolimits }
\def\weight{ \mathop{\rm weight}\nolimits }
\def\Orb{ \mathop{\rm Orb} }
\def\Var{ \mathop{\rm Var}\nolimits }
\def\Cov{ \mathop{\rm Cov}\nolimits }
\def\mean{ \mathop{\bf E}\nolimits }
\def\EE{ \mathop\Ee\nolimits }
\def\PP{ \mathop\Pp\nolimits }
\def\diam{\mathop{\rm diam}\nolimits}
\def\sign{\mathop{\rm sign}\nolimits}
\def\prob{\mathop{\rm Prob}\nolimits}
\def\gap{\mathop{\rm gap}\nolimits}
\def\tto#1{\buildrel #1 \over \longrightarrow}
\def\norm#1{ | #1 | }
\def\scalprod#1#2{ \thsp<#1, \thsp #2>\thsp }
\def\inte#1{\lfloor #1 \rfloor}
\def\ceil#1{\lceil #1 \rceil}
\def\intl{\int\limits}
\outer\def\nproclaim#1 [#2]#3. #4\par{\medbreak \noindent
   \talato(#2){\bf #1 \Thm[#2]#3.\enspace }%
   {\sl #4\par }\ifdim \lastskip <\medskipamount 
   \removelastskip \penalty 55\medskip \fi}
\def\thmm[#1]{#1}
\def\teo[#1]{#1}

%------------------------------ tilde
%
\def\sttilde#1{%
\dimen2=\fontdimen5\textfont0
\setbox0=\hbox{$\mathchar"7E$}
\setbox1=\hbox{$\scriptstyle #1$}
\dimen0=\wd0
\dimen1=\wd1
\advance\dimen1 by -\dimen0
\divide\dimen1 by 2
\vbox{\offinterlineskip%
   \moveright\dimen1 \box0 \kern - \dimen2\box1}
}
\def\ntilde#1{\mathchoice{\widetilde #1}{\widetilde #1}%
   {\sttilde #1}{\sttilde #1}}

%-------------------------------------------------------------------
%
%\sezioniseparate %----------- togliere quando si stampa tutto insieme
%\BOZZA %   
%\let\g=\o %      %------------------ per il Mac 

%%%%%%%%%%%%%%%%Macro
\def\boundpiu#1{\partial{#1}^+}
\def\til#1{\widetilde{#1}}
\def\rect#1#2{{\vcenter{\vbox{\hrule height.3pt
	    \hbox{\vrule width.3pt height#2truecm \kern#1truecm
	    \vrule width.3pt}
	    \hrule height.3pt}}}}
\def\square{\rect{0.15}{0.15}}
%%%%%%%%%%%%%%%%%%%%%%

%%%%%%%%%%%%%%%%%%%%%%%Inizio lavoro
%Sezione 0
\expandafter\ifx\csname sezioniseparate\endcsname\relax%
\input macro \fi
\numsec=0
\numfor=1\numtheo=1\pgn=1
\noindent
\centerline {\bf RENORMALIZATION--GROUP AT CRITICALITY}
\centerline {\bf AND COMPLETE ANALYTICITY OF CONSTRAINED MODELS:}
\centerline {\bf  A NUMERICAL STUDY.}
\vskip 2 truecm
\centerline {Emilio N. M. Cirillo}
\par\noindent
\vskip 0.5 truecm
\centerline {\it Dipartimento di Fisica dell'Universit\`a di Bari and}
\par\noindent
\centerline {\it Istituto Nazionale di Fisica Nucleare, Sezione di Bari.}
\par\noindent
\centerline {\it V. Amendola 173, I--70126 Bari, Italy.}
\par\noindent
\centerline {\rm E--mail: cirillo@axpba0.ba.infn.it}
\vskip 1 truecm
\centerline {Enzo Olivieri}\par\noindent
\vskip 0.5 truecm
\centerline {\it Dipartimento di Matematica - II Universit\`a di Roma}
\par\noindent
\centerline {\it Tor Vergata - Via della Ricerca Scientifica - 00173 ROMA - 
Italy.}
\par\noindent
\centerline {\rm E--mail: olivieri@mat.utovrm.it}
\vskip 1.5 truecm
\centerline {\bf Abstract}
\vskip 0.5 truecm
\centerline{
\vbox{
\hsize=14truecm
\baselineskip 0.4cm
We study the majority rule transformation applied to the Gibbs measure for the
2--D Ising model at the critical point.
The aim is to show that the renormalized hamiltonian is well defined in the 
sense that the renormalized measure is Gibbsian.
We analyze the validity of Dobrushin--Shlosman Uniqueness (DSU)
finite--size condition for the ``constrained models" corresponding to different
configurations of the ``image" system.
It is known that DSU implies, in our 2--D case, complete analyticity from which,
as it has been recently shown by Haller and Kennedy, Gibbsianness follows.
We introduce a Monte Carlo algorithm to compute an upper bound to Vasserstein
distance (appearing in DSU) between finite volume Gibbs measures with different
boundary conditions. We get strong numerical evidence that indeed DSU 
condition is verified for a large enough volume $V$ for all constrained models.
}}

%Sezione 1
\vfill\eject
\expandafter\ifx\csname sezioniseparate\endcsname\relax%
\input macro \fi
\numsec=1
\numfor=1\numtheo=1\pgn=1
\noindent
{\bf 1. Introduction.}
\smallno
\par
In the recent few years many efforts have been devoted to the problem of
a correct definition, on rigorous grounds, of various real--space
renormalization--group maps.
\par 
The main question is whether or not a measure
$$
\n = T_b \m \Eq (1.1)
$$
arising from the application of a renormalization group transformation 
(RGT) $T_b$, defined ``on scale $b$", to the Gibbs measure $\m$ is Gibbsian. 
In other words we ask ourselves whether or not $\n$ is the Gibbs measure
corresponding to a finite--norm translationally invariant potential so that
the ``renormalized hamiltonian" is well defined.\par
To be concrete let us suppose that $\m = \m_{\b,h}$ is the Gibbs measure
describing 2--D Ising model at inverse temperature $\b$ and external
magnetic field $h \neq 0$.
Moreover we assume that our RGT can be expressed as: 
$$
\n(\s') = \sum _{\s} T_b (\s',\s) \m_{\b,h} (\s) \Eq (1.2)
$$
where $T_b (\s',\s)$ is a normalized non--negative kernel. 
The system described in terms of the $\s$ variables by the original measure 
$\m$ is called ``object system". The $\s'$'s are the
``block variables" of the ``image system" described by the renormalized
measure $\n$.
\par 
We can think of the transformation $T_b$ as  directly acting
at infinite volume or we can consider a finite volume version and
subsequently try to perform the thermodynamic limit.
We refer to the basic reference [EFS] for a clear and complete description of 
the general set--up of renormalization maps from the point of view of rigorous
statistical mechanics.
\par
The above mentioned pathological behaviour (non--Gibbsiannes of $\n$)
can be a consequence of the violation of a necessary condition for
Gibbsiannes called {\it quasi--locality} (see [Ko], [EFS]).
It is a continuity property of the finite volume conditional probabilities of 
$\n$ which, roughly speaking, says that they 
are almost independent  
of very far away conditioning spins.\par
In [Ko] it is shown that a sort of converse statement holds true; namely:
quasilocality + nonnulness (uniform positivity of conditional probabilities) of
a stochastic field implies  Gibbsianness but only in the sense of
 the existence of a finite norm but in
general not translationally invariant potential associated to $\n$. 
The construction of the potential in Kozlov's proof is somehow artificial: it
involves reordering of a semi--convergent sum. To get a
translationally invariant finite norm potential one needs some  additional
stronger assumptions  on how weakly the conditional probabilities of $\n$ 
depend on far away conditioning spins. 
We refer to [BMO] for a more detailed discussion on this point.
\par
In some situations (see, for instance [C], [HK]) 
it is possible to use much stronger
methods, based on  cluster expansion, to compute renormalized potentials, 
showing finiteness of their norm. 
\par
In many interesting
examples (see [E1], [EFS], [EFK]) violation of quasi--locality and consequently
non--Gibbsiannes of the renormalized measure $\n$ is a direct consequence of 
the appearance of a first order phase transition for the original (object) 
system described by $\m$ conditioned to some particular configuration of the 
image system. More precisely given a block configuration $\s'$ let us consider 
the probability measure on the original spin variables given by:
$$\m_{\s'} (\s) \; = \; { T_b (\s',\s) \m(\s) 
\over \sum _{\eta}  T_b (\s',\eta) \m(\eta)}$$
It defines the ``constrained" model corresponding to $\s'$
(which here plays the role of an external parameter).
\par
For some particular $\s'$ it may happen that the corresponding measure 
$\m_{\s'}(\s)$ exhibits long range order. See also [GP], [I] where this 
mechanism was first pointed out. 
\par
One can ask himself about ``robustness" of the pathology of non--Gibbsianness.
There are examples (see [MO4]) in which, even though the measure
$\n = T_b \m_{\b,h}$ is not Gibbsian, one has that with the same $\b, h$, by 
choosing $b' > b$ sufficiently large, the measure $\n' = T_{b'} \m_{\b,h}$ is
Gibbsian. Alternatively one can think to iterate the map and, even though 
after the
first step the resulting renormalized measure is not Gibbsian, it may happen,
after a sufficiently large number of iterations that one gets back to the set 
of Gibbsian measures. This is often related to the fact that, given suitable 
values of the parameters $\b, h$ (near the coexistence line 
$h = 0, \b > \b_c$), on a
suitable scale $b$ some constrained model can undergo a phase transition 
(somehow related to the phase transition of the object system); whereas given 
the same $h,\b$, for sufficiently large scale $b$ any constrained model is in 
the one--phase region.
\par
Another notion of robustness of the pathology (see [LV], [MO5]) refers to see 
whether or not it survives after application of a decimation transformation 
(see [EFS]); this can be relevant since decimation  transformation does not 
change the thermodynamic functions or the long range correlations.
\par
Finally we want to say that weaker notions of Gibbsiannes of a renormalized
measure $\n$ can also be considered. The usual notion of Gibbsiannes requires a
control of quasi--locality of $\n$ {\it uniform} in $\s'$. It may happen 
hat the particular $\s'$ responsible for the pathology is highly non--typical 
w.r.t. $\n$. It appears plausible to ask for quasi--locality only for 
$\n$--almost all configurations $\s'$ (see [D2], [FP], [Ma]).
We refer to [E2] for a nice up--to--date review of all the above problematic.
\par 
Now we want to  notice that in many examples it happens that 
even though the object system is well inside the one--phase region, 
nonetheless for some particular block
configuration $\s'$ the corresponding constrained model undergoes a first 
order phase transition.  Conversely there are many indications that if the 
constrained models are in the weak coupling regime then Gibbsianness of the 
renormalized measure follows. Recently Haller and Kennedy gave very 
interesting new rigorous results in this direction. They proved, under very 
general hypotheses, that if {\it all} constrained models are uniformly 
completely analytical (see [DS2], [DS3]) then the renormalized measure is 
Gibbsian with a finite norm potential which can be computed via a convergent 
cluster expansion.
\par 
Let us give now the example of the Block--Averaging Transformation (BAT).
Suppose to partition $\bZ^2$ into squared blocks $B_i$ of side $2$. In this 
case the new measure $\n$ is obtained by assigning to each block $B_i$ an 
integer value $m_i \in \{-4,-2,0,+2,+4\}$ and by computing the probability, 
w.r.t. the original Gibbs measure $\m_{\b,h}$ of the event:
$\sum_{x \in B_i} \s_x = m_i$. Then in this case we have:
$$
T_b(m,\s) \; =\left\{\eqalign{ 
1& \;\;\;\;\hbox {if}\;\; \sum _{x \in B_i} \s_x = m_i\;\;\forall i\cr
0& \;\;\;\;\hbox {otherwise}\cr}\right.
\;\; .
$$
In this case a constrained model is a ``multicanonical" Ising model namely an
Ising model subject to the constraint of having, for every $i$, magnetization 
$m_i$ in the block $B_i$. It has been shown in [EFK] that for BAT 
transformation the constrained model corresponding to 
$m_i = 0 \; \forall \; i$, undergoes a first
order phase transition at low enough temperature and that this implies
violation of quasi--locality and then non--Gibbsianness of the renormalized
measure $\n$. Notice that  for any constrained model with given $\{m_i\}$ the 
value of the external magnetic field $h$ is totally irrelevant. 
On the other hand, for $h$ very large one can prove, by standard methods, 
absence of phase transition for the original model  
in the strongest possible sense: complete analyticity in the
strong Dobrushin--Shlosman sense holds true in this case. This, as
it has been shown in [EFS] gives an example of non--Gibbsiannes of a measure 
$\n$ arising from the application of a renormalization map to a measure $\m$
corresponding to the very weak coupling region. 
We remark that, as it has been shown in [EFS], this non--Gibbsiannes is robust
w.r.t. the choice of the scale $b$ (or w.r.t iteration) whereas for large $h$  
it can be eliminated by applying one decimation transformation (see [MO5]).
\par
Other interesting examples have been found (see [EFK], [E1], [E2]).
\par 
We want to stress that, in
general, it is not sufficient to control that one single constrained model is 
in the one--phase region to imply Gibbsianness of the renormalized measure.
In [CG] and in [BMO] for the BAT transformation it was suggested that the fact 
that the constrained model with $m_i =0$ was in the high temperature phase 
could be sufficient to imply the existence of a finite norm renormalized 
potential. Recently A. van Enter showed with an example that this believe is 
not sufficiently justified and some extra arguments related to the specific 
nature of BAT transformation are needed to imply Gibbsiannes from absence of 
phase transitions for $\{m_i=0\}$ constrained model (see [E1], [E2]).
\par
Thus, in general, the moral is that what is relevant
for Gibbsiannes of $\n$ are the (intermediate) constrained models; we repeat 
that it can be sufficient that even only one constrained model undergoes a 
phase transition with long--range order (despite of the possible very weak 
coupling regime of the object system) to imply non--Gibbsiannes of $\n$; 
whereas, in general, absence of phase transition in a very strong sense is 
needed {\it for all constrained models} to imply Gibbsiannes of $\n$  in the 
strong, cluster expansion, sense as it has been shown in [HK].
\par
In the present paper we will analyze a particular RGT: the majority rule
transformation applied to the 2D critical Ising model. It will be precisely 
defined in next Section 2. This transformation, in the same situation of 
criticality, has been studied by Tom Kennedy in [K2]. The author establishes 
some rigorous results reducing absence of phase transition for some 
particularly relevant constrained models to the verification of some suitable 
``finite size conditions" introduced in [K1]. It is an ``almost computer 
assisted proof" of absence of phase transition for these constrained models.
\par
In the present paper we inquire for the validity, in principle for every 
possible constrained model, of a finite size condition: the 
Dobrushin--Shlosman uniqueness condition (DSU, see Section 2 below) which 
implies (as it will be explained in Section 2) complete analyticity and then, 
using the results of [HK], Gibbsianness of $\n$. 
\par
Strictly speaking the proof of  [HK] does not directly apply to our case since
it requires the condition that the kernel $T(\s',\s)$ is strictly positive for 
every $\s',\s$. Probably this is only a technical restriction that can be 
removed ([K3]). In any case in [HK] the authors claim that for the majority 
rule they are able to obtain an equivalent system with $T(\s',\s) > 0$ by 
first summing out some spins in the original system.
\par 
Our results and their strength are, in a sense, complementary to the ones of 
[K2]. Our study will be numerical but, similarly to [K2] not only in the sense 
of ``traditional" Monte Carlo simulations. Rather, for each constrained model, 
we will try to measure by a computer a quantity appearing in DSU such that if  
we could rigorously prove that it is strictly less than one then we could 
deduce from some  theorems strong properties typical of the one--phase region, 
for arbitrarily large and even infinite systems. 
\par
In [K2] some constrained models were analyzed in terms of a finite size 
condition easier to be satisfied than DSU, for which the author could also 
have provided a computation based on interval arithmetics suited for a 
computer assisted proof. This finite size condition of reference [K1] is not 
sufficient to imply complete analyticity.
In the present paper, as we said before, we try to verify for every 
constrained model DSU condition; but, as it will appear clear in Section 5 we 
can, with nowadays machines, only perform a Monte Carlo calculation.
For many reasons we cannot, at the moment, hope to improve our calculations to 
get a complete control and possibly a computer assisted proof.
We will explain in Section 6, devoted to the conclusions, in what sense our 
results can be considered satisfactory.
\par
The DSU condition involves the calculation of the so--called Vasserstein 
distance between two Gibbs measures in a finite volume with boundary 
conditions differing only in one conditioning site. In a recent paper ([BMO]) 
the authors, in the context of BAT transformation, for one particularly 
relevant constrained model ($\{m_i\} = 0 \; \forall \; i$)
tried to verify the same DSU condition but they were only able to provide  a
(numerical) lower bound for the concerned Vasserstein distance. The reason was 
that a lower bound (see \eqf(2.15')) in terms of total variation distance  
can be found involving thermal averages;  thus this lower bound  is well 
adapted to be studied by Monte Carlo methods but, on the other hand, it is
only able to give some indications on the validity of the true condition and 
this since it appears reasonable to expect that it is a good lower estimate;  
strictly speaking it is only useful to {\it disprove} the condition.
\par
In the present paper we present a Monte Carlo algorithm, inspired by the 
``surgery method" introduced by Dobrushin and Shlosman (see [DS1], [DS2]) that 
we call {\it dynamical surgery}. It provides  an upper bound to the 
Vasserstein distance and so it goes into the correct direction to {\it prove} 
DSU condition.
\par
Our numerical results strongly suggest that indeed DSU condition is satisfied 
in our present situation for all constrained models and this, as we said 
before implies Gibbsiannes of renormalized measure.
\par
The paper is organized as follows: in Section 2 we define in detail our 
majority rule transformation and the constrained models. In Section 3 by using 
``conventional" Monte Carlo methods we provide rough estimates of the critical
temperatures of some particularly relevant constrained models. In Section 4 we 
introduce our algorithm. In Section 5 we give our main numerical results. In 
Section 6 we give the conclusions. In Appendix A we present the computation of 
the best joint representation of two measures which is used in our algorithm.
\bigno

%Sezione 2
\vfill\eject
\expandafter\ifx\csname sezioniseparate\endcsname\relax%
\input macro \fi
\numsec=2
\numfor=1\numtheo=1\pgn=1
\noindent
{\bf 2. The majority rule transformation and the constrained models.}
\smallno
We will consider the usual (ferromagnetic nearest neighbors interaction)
2D Ising model, with zero external field, on a finite square with even 
side: $\L:=\{1,...,2L\}^2\sset\ZZ$ with $L\in\bN^*$. We denote by 
$\s\in\O_{\L}:=\{-1,+1\}^{\L}$ a configuration of the system in $\L$ and
by $\t\in\{-1,+1\}^{\boundpiu{\L}}$ a boundary condition, namely a 
configuration in $\boundpiu{\L}$ defined by
$$\boundpiu{\L}:=\{i\in\ZZ\setminus\L: \exists j\in\L:\; i\; {\rm and}\; j\; 
{\rm are\; nearest\; neighbors}\}\;\; . \Eq(boundary)$$
By $\s_i\in\{-1,+1\}$ and $\t_j\in\{-1,+1\}$ we denote the spin
variables  on the site $i\in\L$ and $j\in\boundpiu{\L}$.
It is also convenient to think of $\t$ as an extended configuration in 
$\{-1,+1\}^{\ZZ\setminus\L}$.
\par
The energy associated to $\s\in\O_{\L}$ with $\t$ boundary condition 
outside $\L$ and zero external magnetic field is given by
$$H^{\t}_{\L}(\s):=-\sum_{<i,j>\atop i,j\in\L}\s_i \s_j -\sum_{<i,j>\atop
i\in\L,j\in\boundpiu{\L}} \s_i\t_j
\;\;\;\forall\s\in\O_{\L}\; ,\Eq(ising)$$
where the first sum runs over all pairs of nearest neighbors sites in 
$\L$, while the second sum runs over all pairs $<i,j>$ of nearest 
neighbors sites such that $i\in\L$ and $j\in\boundpiu{\L}$.
The Gibbs measure describing the equilibrium properties of the system at 
the inverse temperature $\b$ is denoted by 
$\m_{\b,\L}^{\t}(\s)\;\forall\s\in\O_{\L}$ and is given by
$$\m_{\b,\L}^{\t}(\s):={e^{-\b H^{\t}_{\L}(\s)}\over 
\sum_{\h\in\O_{\L}} e^{-\b H^{\t}_{\L}(\h)}}\;\; 
\forall\s\in\O_{\L}\; .\Eq(isingmeas)$$
\par
In the following four steps we give the precise definition of the 
``Majority Rule"  (on scale 2) transformation.
\medno
1. $\forall x,y\in\bZ$ let $B_{(x,y)}$ denote the $2\times 2$ block 
whose center has coordinates: 
\par\noindent $\left(2x-{1\over 2},2y-{1\over 2}\right)$; 
the collection of all  blocks $B_{(x,y)}\;\forall x,y\in\bZ$ 
gives rise to a partition of the lattice $\bZ^2$. If we restrict ourselves 
to pairs $(x,y)\in\{1,...,L\}^2$ we get a partition of our box $\L$.
Given $\s\in\O_{\L}$ we denote by $\s^1_{(x,y)},...,\s^4_{(x,y)}$ the 
four spins corresponding to the four sites of the block 
$B_{(x,y)}$ and we define 
$m_{(x,y)}:=\sum_{i=1}^4\s^i_{(x,y)}$; we suppose the four 
spins $\s^i_{(x,y)}\;\forall i=1,...,4$ ordered in lexicographic way.
\smallno
2. We define the new lattice $\L'$ by collecting the centers of 
all  $B_{(x,y)}$ blocks and by rescaling the lattice spacing by a factor 
two; the site of $\L'$, which is the center of the block $B_{(x,y)}$, will be 
simply denoted by the pair $(x,y)$.
\smallno
3. On each site $(x,y)\in\L'$ we define the {\it renormalized spin}
$\s'_{(x,y)}\in\{-1,+1\}$ and 
we consider the space $\O'_{\L'}:=\{-1,+1\}^{\L'}$. We define the kernel 
$K: (\s,\s')\in\O_{\L}\times\O'_{\L'}\rightarrow K(\s,\s')\in\{0,1\}$ as 
follows
$$K(\s,\s'):=\left\{
\eqalign{
0&\; {\rm if}\;\exists\; (x,y)\in\L' :\; m_{(x,y)}\not= 0 \;
	{\rm and}\; m_{(x,y)}\cdot\s'_{(x,y)}<0\cr 
0&\; {\rm if}\;\exists\; (x,y)\in\L' :\; m_{(x,y)}=0 \;{\rm and}\;
		    \s^1_{(x,y)}\cdot\s'_{(x,y)}<0\cr
1&\; {\rm otherwise}\cr}
\right.\;\; .\Eq(kernel)$$
\smallno
4. The {\it Majority Rule Transformation} is the transformation which 
maps the ``object" model $(\L,\O_{\L},\m^{\t}_{\b,\L}(\s))$ onto the 
``image" model $(\L',\O'_{\L'},\m'^{\t}_{\b,\L'}(\s'))$, where 
$$\m'^{\t}_{\b,\L'}(\s'):={\sum_{\s\in\O_{\L}} K(\s,\s')\m^{\t}_{\b,\L}(\s)
\over\sum_{\h'\in\O'_{\L'}}\sum_{\s\in\O_{\L}} K(\s,\h')
\m^{\t}_{\b,\L}(\s)}\;\;\;\;
\forall\s'\in\O'_{\L'}\; .\Eq(objmeas)$$
Notice that we could have used the notation $\O_{\L'}$ in place of $\O'_{\L'}$
since here, for the majority rule transformation, contrary to other 
transformations like BAT, the single renormalized spin variable still takes 
values in $\{-1,+1\}$.
\medno
\par
This transformation is well known in physics literature and it has been widely 
used to investigate the properties of many spin models 
(see, e. g., [NL] and references therein).
\par
A very 
important role in our discussion will be  played by the ``constrained models":
given  $\s'\in\O'_{\L'}$ 
we call {\it constrained model corresponding to} $\s'$, and we denote it by 
$\cI^{\t}_{\b,\s'}$, the model 
$(\L,\O_{\L},\m^{\t}_{\b,\L,\s'}(\s))$, where
$$\m^{\t}_{\b,\L,\s'}(\s):={K(\s,\s')\m^{\t}_{\b,\L}(\s)\over
\sum_{\h\in\O_{\L}} K(\h,\s')\m^{\t}_{\b,\L}(\h)}\Eq(intermeas)$$
is a probability measure on $\O_{\L}$. Due to the fact that 
$K(\s,\s')\in\{0,1\}$ we have that, $\forall\s\in\O_{\L}$ and
$\forall\s'\in\O'_{\L'}$ the  constrained model $\cI^{\t}_{\b,\s'}$ can be 
seen as the model  $$\left(\L,\O_{\L,\s'},\m^{\t}_{\b,\L,\s'}(\s)=
{e^{-\b H^{\t}_{\L}(\s)}\over
\sum_{\h\in\O_{\L,\s'}}e^{-\b H^{\t}_{\L}(\h)}}\right)\; ,\Eq(constr-model)$$ 
where we have introduced the {\it constrained configuration space}
$$\O_{\L,\s'}:=\{\s\in\O_{\L}:\;\forall(x,y)\in\L':$$
$$(m_{(x,y)}\not= 0 \; {\rm and}\; m_{(x,y)}\cdot\s'_{(x,y)}>0)\; 
{\rm or} \;(m_{(x,y)}=0 \;{\rm and}\;\s^1_{(x,y)}\cdot\s'_{(x,y)}>0)
\}\;\; .\Eq(constrain)$$
In other words we can say that the constrained 
model $\cI^{\t}_{\b,\s'}$ is a model defined on 
the original lattice $\L$, with the same hamiltonian $H^{\t}_{\L}(\s)$ 
as the object model, but with configuration space $\O_{\L,\s'}$.
\par
The microscopic states of the constrained models can be characterized by 
means of a suitable {\it block--variable}. In the original Ising model there 
are $2^4=16$ allowed configurations in each block $B_{(x,y)}$. We partition 
these block--configurations into two disjoint classes $C_+$ and $C_-$ 
as follows:
\medno
{$\bullet$} block--configurations belonging to class $C_+$
$$\left[\matrix{-&+\cr +&+\cr}\right]\;
\left[\matrix{+&-\cr -&+\cr}\right]\;
\left[\matrix{+&-\cr +&-\cr}\right]\;
\left[\matrix{+&-\cr +&+\cr}\right]\;
\left[\matrix{+&+\cr -&-\cr}\right]\;
\left[\matrix{+&+\cr -&+\cr}\right]\;
\left[\matrix{+&+\cr +&-\cr}\right]\;
\left[\matrix{+&+\cr +&+\cr}\right]\Eq(class+)$$
\medno
{$\bullet$} block--configurations belonging to class $C_-$
$$\left[\matrix{-&-\cr -&-\cr}\right]\;
\left[\matrix{-&-\cr -&+\cr}\right]\;
\left[\matrix{-&-\cr +&-\cr}\right]\;
\left[\matrix{-&-\cr +&+\cr}\right]\;
\left[\matrix{-&+\cr -&-\cr}\right]\;
\left[\matrix{-&+\cr -&+\cr}\right]\;
\left[\matrix{-&+\cr +&-\cr}\right]\;
\left[\matrix{+&-\cr -&-\cr}\right]\Eq(class-)$$
\medno
For any constrained model $\cI^{\t}_{\b,\s'}$ we allow, in the block 
$B_{(x,y)}$, only the 
configurations in the class $C_{\sign\s'_{(x,y)}}$; in order to 
classify these block--configurations we introduce the block--variable
$S_{(x,y)}\in\{1,2,...,8\}$: to the values 
$S_{(x,y)}=1,...,8$ correspond, respectively, the eight block--configurations 
in \equ(class+) if $\s'_{(x,y)}=+1$, the eight block--configurations 
in \equ(class-) if $\s'_{(x,y)}=-1$. 
\bigno
{\bf Warning.}
\par\noindent
Here, and in what follows, we use the (non--conventional)  expression 
{\it block--variable} referring to the variable $S_{(x,y)}$ taking values in 
the set $\{1,2,...,8\}$; the spin variables $\s'_{(x,y)}$, defined on the 
renormalized lattice $\L'$ and taking values in $\{-1,+1\}$, will be 
sometimes called {\it renormalized  variables}.
\bigno
\par
Obviously, given any constrained model $\cI^{\t}_{\b,\s'}$, one and only 
one configuration $S\in\til{\O}_{\L'}:=\{1,2,...,8\}^{\L'}$ can be associated 
to any $\s\in\O_{\L,\s'}$, and viceversa. 
Hence, each state of the model $\cI^{\t}_{\b,\s'}$ can be represented by a 
collection $S\in\til{\O}_{\L'}$ of block--variables $S_{(x,y)}$, but we 
recall that the ``meaning" (in terms of the original spin variables)
of each block--variable $S_{(x,y)}$ depends on the sign of $\s'_{(x,y)}$.
\par
Due to the bijection between $\O_{\L,\s'}$ and $\til{\O}_{\L'}$ the 
hamiltonian $H^{\t}_{\L}(\s)$ of the model $\cI^{\t}_{\b,\s'}$
can be thought, for any $\s\in\O_{\L,\s'}$, as a function 
$H^{\cT}_{\L',\s',\t'}(S)$ of 
the block--variables configuration $S\in\til{\O}_{\L'}$; 
here by $\t'$ we mean a configuration of the renormalized variables in 
$\boundpiu{\L'}$, that is 
$\t'\in\O'_{\boundpiu{\L'}}:=\{-1,+1\}^{\boundpiu{\L'}}$, where
$$\boundpiu{\L'}:=\{(x,y)\in\ZZ: 0\le x,y\le L+1,\; x\not= y,\; 
                   (x,y)\not\in\L'\}\; ; \Eq(boundarynew)$$
by
$\cT\in\til{\O}_{\boundpiu{\L'}}:=\{1,...,8\}^{\boundpiu{\L'}}$ we mean
the boundary condition expressed in terms of the block--variables in the set
$\boundpiu{\L'}$.
Notice that given the original $\t$ and $\t'$ in general $\cT$ is not uniquely
determined. 
\par
We remark  
that a block--variable in $B_{(x,y)}$, for $(x,y)\in\L'\cup\boundpiu{\L'}$, 
is completely ``meaningless" if the correspondent renormalized variable has 
not been specified. Finally, one can say that the constrained model  
\equ(constr-model) is equivalent to the model defined on the lattice $\L'$, 
with configuration space $\til{\O}_{\L'}$ and equilibrium measure given 
by
$$\m^{\cT}_{\b,\L',\s',\t'}(S):=
{e^{-\b H^{\cT}_{\L',\s',\t'}(S)}\over
\sum_{\Y\in \til{\O}_{\L'}} e^{-\b H^{\cT}_{\L',\s',\t'}(\Y)}}
\;\;\;\;\forall S\in\til{\O}_{\L'}\;\; .\Eq(block-meas)
$$
\par
Now, we want to study the above defined constrained models by means of the 
{\it finite size Dobrushin--Shlosman uniqueness condition} (DSU);  to 
introduce DSU 
we need some  
definitions.\par
Let us consider two measures $\m_1$ and $\m_2$ on a finite set 
$Y$; let $\r(\cdot,\cdot)$ be a metrics on $Y$ and denote by 
$\cK(\m_1,\m_2)$ the set of joint representations of $\m_1$ and $\m_2$, 
namely the set of all  measures $\m$ on the Cartesian product $Y\times Y$ 
such that 
$$\sum_{y\in B,y'\in Y} \m(y,y')=\m_1(B)\;\;\;\; {\rm and}\;\;
\sum_{y\in Y,y'\in B} \m(y,y')=\m_2(B)
\;\;\;\forall B\subset Y\;\; .$$
We set:
$${ Var}(\m_1,\m_2):={1\over 2}\sum_{y\in Y}|\m_1(y)-\m_2(y)|\;\;
\Eq(variation)$$
and, given a metrics $\r$ on $Y$
$$\cD_{\r}(\m_1,\m_2):=\inf_{\m\in\cK(\m_1,\m_2)}
      \sum_{y,y'\in Y} \r(y,y')\cdot \m(y,y')
\;\; ;\Eq(vasser)$$
${ Var}(\m_1,\m_2)$ and $\cD_{\r}(\m_1,\m_2)$ are respectively
called {\it total variation distance} and {\it Vasserstein distance with 
respect to $\r$} between the two measures $\m_1$ and $\m_2$.
\par
Let us consider a spin system on $\bZ^d$ with single spin space $\cS$ and 
range--one interaction  (generalization to arbitrary finite range is trivial); 
we denote by $\eta_i$ the spin variable  associated to the site $i\in\bZ^d$. 
\par
For any finite set $V\subset\bZ^d$ we denote by $\boundpiu{V}$ the set of 
points outside $V$ whose spins interact with the spins inside $V$ and by 
$\eta_{V}^{}\in\cS^{V}$ a spin configuration on $V$. Given the boundary 
condition $\xi\in\cS^{V^c}$, we denote by $\m_{V}^{\xi}$ the Gibbs measure 
in $V$ with boundary conditions $\xi$ outside $V$. Given a metrics $\r$ 
on $\cS$ we associate to it a metrics $\r_{V}^{}$ on $\cS^{V}$ defined as 
follows
$$\r_{V}^{}(\eta_{V}^{},\eta'_{V}):=\sum_{i\in V}\r(\eta_i^{},\eta'_i)
\;\;\;\;\forall\eta_{V}^{},\eta'_{V}\in\cS^{V}\;\; .$$
\par
We say that condition $DSU_{\r}(V,\d)$ (the Dobrushin--Shlosman
Uniqueness condition in $V$ with respect to the metrics $\r$ and uniqueness
parameter $\d$) is satisfied if and only if
$\exists$ a finite set $V\subset\bZ^d$ and $\exists\d >0$ such that:  
$\forall j\in\boundpiu{V}$ $\exists \a_j>0$ such that for any couple of 
boundary conditions $\xi,\xi'\in\cS^{V^c}$ with $\xi_i=\xi'_i\;\forall 
i\not= j$ one has
$$\cD_{\r _V}(\m_{V}^{\xi},\m_{V}^{\xi'})\le\a_j\r(\xi_j^{},\xi'_j)\;\; ,$$
and
$$\sum_{j\in\boundpiu{V}}\a_j\le\d |V|\;\; .$$
\par
Let us consider the metrics on the single site variable given by:
$$\til{\r} (\h,\h'):=\left\{ \eqalign{
1&\;\;\; {\rm iff}\; \h\not=\h'\cr
0&\;\;\; {\rm otherwise}\cr}\right.
\;\; ;\Eq(dist-singo)$$
in this case condition $DSU_{\tilde \r}(V,\d)$ will be simply denoted by 
$DSU(V,\d)$.
\par
We observe that, in this case
$$\cD_{\r_V^{}}(\m_{V}^{\xi},\m_{V}^{\xi'}) \ge\sum _{i \in V}
\cD_{\tilde \r}(\m_{i}^{\xi},\m_{i}^{\xi'})
= \sum _{i \in V}  Var ( \m_{i}^{\xi},\m_{i}^{\xi'})
\Eq (2.15')
$$ 
as it easily follows from \equ (vasser) and Proposition A1 in the Appendix.
\par
In order to describe Dobrushin--Shlosman's results based on $DSU_{\r}(V,\d)$
condition we need some more definitions.
We are going to introduce  two kinds  of  mixing conditions,       
of a priori different strength, for Gibbs measures in a finite volume $\L$.
The first ones are  weak mixing (WM) conditions saying that the influence of a
local change in conditioning spins decays exponentially fast with the distance
from the boundary $\partial \L$; the second kind, the one of strong mixing
(SM) conditions, corresponds to the case where the influence of a change of
a conditioning spin $x$ decays exponentially fast with the distance from $x$. 
We refer to [MO1], [MO2] for a critical discussion of these different notions. 
It has been shown with some examples (see [Sh]) that it may happen that WM is
satisfied whereas the corresponding SM is not. Since we are speaking of
finite volume mixing condition we have to make explicit in the notation
the constants referring to the concerned exponential decay. Of course a
particularly interesting case is when we have these mixing conditions in a
class of arbitrarily large volumes $\L$ with uniform constants.
We refer again to [MO1], [MO2] for a discussion of these point. It turns out 
that a crucial aspect is the class of volumes that we are considering. In
particular in the Dobrushin--Shlosman theory of complete analyticity
(see [DS2], [DS3]) arbitrary shapes were considered whereas in the approach 
developed in [O], [OP], [MO2] only sufficiently regular domains were involved.
\par
We say that a Gibbs measure $\mu_\Lambda^\tau $ on $\Omega_\Lambda$
satisfies a {\it strong mixing} condition with constants $C, \gamma$
if for every subset $\Delta \subset \Lambda $:
$$ \sup_{\tau,\tau^{(y)} \in \Omega _{\Lambda^c}}
Var(\mu_{\Lambda ,  \Delta}^\tau\ \ \mu_{\Lambda ,
\Delta}^{\tau^{(y)}})\leq C  e^{-\gamma \hbox{dist}(\Delta, y)}
\Eq(2.3)$$
where\quad $\tau^{(y)}_x = \tau_x$ for $ x\ne y$
and $\mu_{\Lambda ,  \Delta}^\tau$ is the $\mu_{\Lambda }^\tau$--probability
 distribution of the spins in $\D$. We denote this condition by
$ SM(\Lambda, C,\gamma)$.
\par
We say that a Gibbs measure $\mu^\tau_\Lambda$ satisfies a {\it weak 
mixing} condition with constants $C , \gamma$ if for every subset
$\Delta \subset \Lambda $
$$\sup_{\tau, \tau'\in\Omega_{\Lambda ^c}}
\quad Var(\mu^\tau_{\Lambda , \Delta},
\mu^{\tau'}_{\Lambda , \Delta})\leq C\sum_{x\in\Delta ,\,y\in
\partial\Lambda^+ }\hbox{exp}(-\gamma\vert x-y\vert)
\;\; ;\Eq(2.5)$$
we denote this condition by $WM(\Lambda , C, \gamma)$.
\par
\medno
{\bf Theorem DS} (see [DS1])
\smallno
Let $DSU (V, \delta)$ be satisfied for some
$V$ and {\bf $\delta < 1$}; then $\exists\ \ C >
0, \gamma > 0$ such that condition $WM(\Lambda , C, \gamma)$
holds {\it for every $\Lambda $}.
\par
\medno
{\bf Theorem MOS} (see [MOS]) \
\smallno
Let the dimension of the lattice be $d=2$. If there
exist positive constants $C$ and $\g$ such that the Gibbs measure
$\mu_\L^\t$ satisfies the weak mixing condition $WM(\Lambda , C,
\gamma)$ for any finite $\L\subset\Z$, then there exist positive
constants  $C'$ and $\g '$ such that the Gibbs measure
$\mu_{\L}^\t$ satisfies the strong mixing condition $SM(\Lambda ,
C', \gamma')$ for any sufficiently regular domain $\L$ and in particular
for any square $\L_L$ with arbitrary side $L$.
\medno
Here ``sufficiently regular" means ``multiple of a sufficiently large square" 
(see [MO1]).
\par
\bigno
{\bf Remark}
\par\noindent
Actually the conclusion of the above theorem remains true even if we assume
the weak mixing {\it not} for {\it all} finite subsets of $\Z$ but {\it
only} for all subsets of a square $\L_{L_o}$ provided that
$L_o$ is large enough (depending of course on the constants $C$ and
$\gamma$ and on the range of the interaction). 
\bigno
\par 
Then our strategy to show Gibbsianness of the renormalized measure arising
from the application of the Majority Rule transformation is based on the
following chain of implications.
\bigno
\par 
We try to verify $DSU(V,\d)$ for some given $V$ and $\d < 1$ for any 
constrained model. Then if we could apply Theorem DS we would get 
$WM(\Lambda , C, \gamma)$ $\forall \L$ with the same constants 
$C,\g$ for all $\L$ and for all 
constrained model. Subsequently, if we could apply Theorem MOS (by exploiting
two--dimensionality), we would get $SM(\Lambda, C',\gamma')$ for all
sufficiently regular domains $\L$ with the same constants $C',\g'$ for all
these $\L$ and for all constrained model. This would directly imply the 
validity of the conclusions of Theorem 1.1 in [HK]; indeed it immediately 
follows from the proof of Theorem 1.1 in [HK] that the authors could have 
obtained exactly the same result by only  assuming their strong mixing 
hypothesis, (that they express in an equivalent form, valid for Ising--like 
systems, as exponential decay of two--points spin--spin  truncated
correlations) uniformly in the constrained model and in
the boundary conditions {\it only  for all sufficiently regular} (in the
above specified sense) domains  instead of {\it for all} domains. 
\par
Strictly speaking Theorems DS and MOS apply to translationally invariant
situations and our constrained models are not, in general, translationally
invariant. However it is easy to convince oneself that both Theorems extend
in a straightforward way to the non--translationally invariant case provided
we assume spatial uniformity of the bounds appearing in the hypotheses.
In other words we have that it is sufficient to assume the  validity, 
for a given constrained model, of $DSU (V,\d)$ for some $V$ and some
$\d < 1$ {\it uniformly} in the location of $V$ to imply $SM$ 
(for arbitrary sufficiently regular volumes with uniform constants) for the 
same model. But if we are able to show that $DSU (V,\d)$ is verified
for a given $V$ and $ \d <1$ {\it for all} constrained models 
(namely for all $\s'\in \O_{V'}$) this implies, at the same time, 
(using extended versions of Theorem DS and MOS) the validity of SM 
(for arbitrary sufficiently regular volumes with uniform constant)
for all constrained models and then, via Theorem 1.1 in [HK], Gibbsianness of 
the renormalized measure.
\bigno
\par
Now  we start applying the above strategy. We first introduce some specific
definitions. Let us consider a squared volume
$V'\subset\L'$ with side 
$l$ and the corresponding subset $V$, with side $2l$, of the original lattice 
$\L$. Let us consider the metrics on the single block--variable space 
$\{1,...,8\}$ 
defined by
$$\til{\r} (\a,\a'):=\left\{ \eqalign{
1&\;\;\; {\rm iff}\; \a\not=\a'\cr
0&\;\;\; {\rm otherwise}\cr}\right.
\;\;\;\; \forall\a,\a'\in\{1,...,8\}\;\; .\Eq(dist-sing)$$
and the metrics on $\til{\O}_{V'}$ given by
$$ 
\r_{V'}^{} (S,S'):=\sum_{(x,y)\in V'} \til{\r} (S_{(x,y)},S'_{(x,y)})
\;\;\;\;\forall S,S'\in \til{\O}_{V'} \;\; .\Eq(dist-block)$$
\par
Now, given $\s'\in\O'_{V'}$ and $\t'\in\O'_{\boundpiu{V'}}$, we denote by 
$[\cT_1,\cT_2]_{(x,y)}$ a pair of boundary conditions 
$\cT_1$, $\cT_2\in \til{\O}_{\boundpiu{V'}}$
such that $\cT_{1_{(x',y')}}=
\cT_{2_{(x',y')}}\;\forall (x',y')\in\boundpiu{V'}\setminus\{(x,y)\}$
and we define
$$\cE^{[\cT_1,\cT_2]_{(x,y)}}_{\b,V',\s',\t'}:=
\cD_{\r_{V'}^{}}(\m^{\cT_1}_{\b,V',\s',\t'},
\m^{\cT_2}_{\b,V',\s',\t'}) {|\boundpiu{V'}|\over |V'|} =
\cD_{\r_{V'}^{}}(\m^{\cT_1}_{\b,V',\s',\t'},
\m^{\cT_2}_{\b,V',\s',\t'}) {4\over l}
\;\; ,\Eq(estim-vero)$$
where $\cD_{\r_{V'}^{}}(\m^{\cT_1}_{\b,V',\s',\t'},
\m^{\cT_2}_{\b,V',\s',\t'})$ is the Vasserstein distance between 
two equilibrium measures for the constrained model corresponding to 
$\s',\t'$
which have been obtained by modifying the boundary conditions just in one 
site in $\boundpiu{V'}$. We set:
$$\cE_{\b,V',\s'}:=
\sup_{(x,y)\in\boundpiu{V'}}\;\sup_{[\cT_1,\cT_2]_{(x,y)}} 
\cE^{[\cT_1,\cT_2]_{(x,y)}}_{\b,V',\s',\t'}\;\; ;
\Eq(estim-indep)$$
heuristically we can say that $\cE_{\b,V',\s'}$ measures how much the 
equilibrium of the constrained model is modified if one changes the boundary 
condition in one site, uniformly in the site and in the boundary 
conditions; it can be called {\it uniqueness parameter} (in the sense of DSU
condition) in $V'$ for the constrained model characterized by $\s'$ at inverse
temperature $\b$ \par 
It is a trivial consequence of the definitions that if 
$\cE_{\b,V',\s'}<1$ then the Dobrushin--Shlosman uniqueness condition 
$DSU(V',\d)$ is satisfied for some $\d<1$ for the 
constrained model corresponding to $\s',\t'$. The main aim of this paper is 
to build up a Monte Carlo algorithm in order to estimate 
$\cE_{\b,V',\s'}$ and to show that at 
$\b=\b_c:={1\over 2}\log (1+\sqrt 2)$ (the critical inverse temperature of 
the standard 2D Ising model) there exists a volume $V'$ such that 
for all  possible constrained models $\cE_{\b,V',\s'}<1$, namely:
$$
\cE_{\b,V'}:=\sup_{\s'\in\O'_{V'}} \cE_{\b,V',\s'}\; <1\;\;.
\Eq(estim-mod-indep)
$$
\bigno

%Sezione 3
\vfill\eject
\expandafter\ifx\csname sezioniseparate\endcsname\relax%
\input macro \fi
\numsec=3
\numfor=1\numtheo=1\pgn=1
\noindent
{\bf 3. Rough estimate of critical temperatures.}
\smallno
In the following sections we will see 
that two constrained models (the {\it chessboard} and the {\it striped} model,
see  definition below) are particularly ``dangerous", that is they 
satisfy  DSU condition for volumes larger than 
those needed by the other constrained models.
\par
The {\it chessboard} and the {\it striped} models are the constrained models 
corresponding, respectively, to the two configurations $\cC,\cS\in\O'$ 
defined as follows
$$\cC_{(x,y)}:=\left\{\eqalign{
1&\;\;\; {\rm if}\; x+y\; {\rm is \; even}\cr
-1&\;\;\; {\rm otherwise}}\right.\;\;\;
\cS_{(x,y)}:=\left\{\eqalign{
1&\;\;\; {\rm if}\; y\; {\rm is \; even}\cr
-1&\;\;\; {\rm otherwise}}\right.\; ;\Eq(chess-stri)$$
these two configurations are respectively depicted in Fig.1b and Fig.1c.
\par
We study the equilibrium properties of the chessboard and the striped 
model by means of a standard Monte Carlo procedure based on a suitable Heat 
Bath dynamics. This dynamics is generally used to compute mean values 
(with respect to the equilibrium Gibbs measure) of some observables as 
time averages; this mean value will be denoted by $<\cdot>$.
\par
This ``conventional Monte Carlo" analysis is preliminar to the main 
numerical results of this paper. Here
we are not interested in a large scale simulation nor 
in a precise estimate of critical points and exponents; we just want to 
have a strong evidence of the fact that the critical inverse 
temperature of the two dangerous models is, in both cases, much greater 
than $\b_c$.
\par
By making use of the notation introduced in Section 2 (relatively to the 
constrained models $\cI^{\t}_{\b,\s'}$) 
we describe, now, the discrete time Heat Bath dynamics used in our Monte Carlo 
study. It is given by the Markov chain defined below:
\medno
1. We consider the constrained model corresponding to 
$\s'\in\O'_{\L'}$ and we assume periodic boundary conditions; we 
denote it by $\cI_{\b,\s'}$ and its hamiltonian by $H_{\L',\s'}(S)$ with 
$S\in\til{\O}_{\L'}$.
\smallno
2. At each step we perform a complete updating of all   
$S_{(x,y)}$ block--variables following the lexicographic order; in the 
remaining of this section this will be called a Monte Carlo {\it sweep}.
\smallno
3. For any $(x,y)\in\L'$ the new value $S'_{(x,y)}$ of the 
block--variable $S_{(x,y)}$ is chosen at random according to the Gibbs 
measure in $B_{(x,y)}$ with boundary condition 
$S^c_{(x,y)}:=S|_{(\L'\cup\partial\L'_+)\setminus B_{(x,y)}}$. 
Hence, if we denote by $H_{\s'}(S'_{(x,y)}|S^c_{(x,y)})$ the contribution of 
the block $B_{(x,y)}$ to the energy of the system, the transition probability 
is given by
$$P_{\b,\s'}(S_{(x,y)}\rightarrow S'_{(x,y)}):=
     {\exp[-\b H_{\s'}(S'_{(x,y)}|S^c_{(x,y)})]\over
      \sum_{S''_{(x,y)}=1}^8 \exp[-\b H_{\s'}(S''_{(x,y)}|S^c_{(x,y)})]}
\;\; .\Eq(transition)$$
We remark that in our notation there is'nt any explicit dependence of the 
boundary conditions  ($\cT$--$\t'$), because they are supposed to be periodic.
\medno
\par
In order to estabilish for which value of the inverse temperature $\b$ the 
two models $\cI_{\b,\cC}$ and $\cI_{\b,\cS}$ are critical, we have 
computed the specific heat defined in terms of the equilibrium energy 
fluctuations:
$$C_{\L,\s'}:={\b^2\over 4L^2}(<H^2_{\L',\s'}>-<H_{\L',\s'}>^2)\;\;\;\;
\forall \s'=\cC,\cS
\;\; .\Eq(heatbath)$$
\par
Both models have been studied in the case $L=64$, which means that we 
have considered the two models defined on square lattice containing $128^2$ 
original sites; we have considered smaller values of $L$, 
as well, in order to check the 
finite size behaviour. In the case of model $\cI_{\b,\cC}$ we have performed 
$10^5$ full sweeps of our Monte Carlo algorithm for each value of $\b$, while 
in the case of model $\cI_{\b,\cS}$ $1.5\times 10^5$ sweeps have been 
performed.
\par
In Fig.2 and in Fig.3 we have plotted the specific heat as a function of 
the inverse temperature $\b$ in the case of models $\cI_{\b,\cC}$ and
$\cI_{\b,\cS}$ respectively; these results have been obtained by 
analyzing the Monte Carlo data by means of the ``jackknife" procedure. From 
the pictures it is clear that both inverse critical temperatures, 
$\b_c^{\cC}$ and $\b_c^{\cS}$, are significantly greater than the Ising 
inverse critical temperature $\b_c$. Our rough estimates are the following
ones:
$$\b_c^{\cC}=1.60\pm 0.05\;\;\;\;\b_c^{\cS}\ge 2.2\; .
\Eq(critical)$$
\bigno

%Sezione 4
\vfill\eject
\expandafter\ifx\csname sezioniseparate\endcsname\relax%
\input macro \fi
\numsec=4
\numfor=1\numtheo=1\pgn=1
\noindent
{\bf 4. The algorithm.}
\smallno
Let us consider the volume $V'$ introduced in Section 2, 
$\s'\in\O'_{V'}$, $\t'\in\O'_{\boundpiu{V'}}$,
a pair of boundary conditions $[\cT_1,\cT_2]_{(\bar{x},\bar{y})}$, the two 
equilibrium measures $\m^{\cT_1}_{\b,\L',\s',\t'}$ and 
$\m^{\cT_2}_{\b,\L',\s',\t'}$. Let us denote by 
$\cK^{[\cT_1,\cT_2]_{(\bar{x},\bar{y})}}_{\b,V',\s',\t'}$ the set of joint 
representations of $\m^{\cT_1}_{\b,\L',\s',\t'}$ and 
$\m^{\cT_2}_{\b,\L',\s',\t'}$;
we want to give a numerical estimate of the quantity
$\cE^{[\cT_1,\cT_2]_{(\bar{x},\bar{y})}}_{\b,V',\s',\t'}$ defined in 
\equ(estim-vero). 
\par
A similar problem has been studied in [BMO]: in that
paper, by means of a standard Heat Bath dynamics, a lower bound to 
$\cE^{[\cT_1,\cT_2]_{(\bar{x},\bar{y})}}_{\b,V',\s',\t'}$ has been 
calculated; now, we build up a Monte Carlo algorithm in order to obtain a 
numerical estimate of an upper bound to the same quantity.
\par
First of all we heuristically describe the idea on which the algorithm
is based. In order to exactly
calculate  the Vasserstein distance one should know a joint 
probability measure 
$Q^{*}(\cdot,\cdot)\in\cK^{[\cT_1,\cT_2]_{(\bar{x},\bar{y})}}
_{\b,V',\s',\t'}$ on the space $\til{\O}_{V'}\times\til{\O}_{V'}$ such that
$$\cD_{\r_{V'}^{}}(\m^{\cT_1}_{\b,V',\s',\t'},\m^{\cT_2}_
{\b,V',\s',\t'})=\sum_{S^{(1)},S^{(2)}\in\til{\O}_{V'}}
Q^{*}(S^{(1)},S^{(2)}) \r_{V'}^{}(S^{(1)},S^{(2)})\;\; ,\Eq(Q-star)$$
that is a joint probability measure ``optimizing" the sum in \equ(vasser);
this is a very hard task. On the other hand it is possible to 
calculate the Vasserstein distance between two equilibrium measures, 
relative to two different local boundary conditions, of a single 
block--variable; indeed, in this case, from the definitions given in Section 
2, it turns out that the distance between two 
configurations is $\til{\r} (\a,\a')=1-\d_{\a,\a'}
\;\forall\a,\a'\in\{1,...,8\}$ (see \equ(dist-sing)) and so the Vasserstein 
distance coincides with the total variation distance (see [D1], page 472). In
this case it is also possible to calculate the ``optimizing" joint 
distribution measure. Hence the idea is to build up a dynamics
describing the evolution of two coupled systems, with different boundary
conditions, such that at each step a single block--variable is 
updated according to the ``local optimizing" joint distribution law.
\par
Now we give the detailed definition of the dynamics.
\medno
1. Given $\s'\in\O'_{V'}$, $\t'\in\O'_{\boundpiu{V'}}$ and
a pair of boundary conditions $[\cT_1,\cT_2]_{(\bar{x},\bar{y})}$, we want 
to describe the evolution of the pair of copies of the constrained model 
corresponding to $\s'\in\O'_{V'}$ and $\t'\in\{-1,+1\}^{\boundpiu{V'}}$
with boundary conditions respectively given by $\cT_1$ and $\cT_2$.
\smallno
2. We consider the Markov Chain $X_t:=(S^{(1)}(t),S^{(2)}(t))\;\forall 
t\in\bN$ where $S^{(1)}(t)$ and $S^{(2)}(t)$ are the configurations of the 
two copies of the system at the time $t$. The two processes $S^{(1)}(t)$ 
and $S^{(2)}(t)$ are called {\it marginal chains}, while $X_t$ is 
called {\it joint chain}.
\smallno
3. At each instant $t$ one and only one site $(x,y)\in V'$ is taken into 
account; all  sites $(x,y)\in V'$ are visited in lexicographic order, 
hence in an interval of time $\D t=|V'|$ all  sites $(x,y)\in V'$ are 
considered: this is called a {\it sweep} (or, sometimes, {\it full sweep}) 
of the dynamics.
\smallno
4. Let us consider the site $(x,y)$ which is updated at the time $t$; we 
introduce the following notation: 
$q^{\cT_i}[S^{(i)}_{(x,y)}|S^{(i)}_{V'\setminus\{(x,y)\}}(t-1)]$
with $i=1,2$ is the equilibrium Gibbs probability measure for the 
block--variable in $(x,y)$  conditioned to the values of the 
block--variables in $V'\setminus\{(x,y)\}$ at time $t-1$; we denote by 
$\cK_{(x,y)}(t)$ the set of joint representations of 
$$
q^{\cT_1}[S^{(1)}_{(x,y)}|S^{(1)}_{V'\setminus\{(x,y)\}}(t-1)]
\;\; {\rm and}\;\;  
q^{\cT_2}[S^{(2)}_{(x,y)}|S^{(2)}_{V'\setminus\{(x,y)\}}(t-1)]
\Eq(marginali)
$$
and, finally, by
$$q^{*\; \cT_1,\cT_2}[S^{(1)}_{(x,y)},S^{(2)}_{(x,y)}
|S^{(1)}_{V'\setminus\{(x,y)\}}(t-1)),S^{(2)}_{V'\setminus\{(x,y)\}}(t-1)]
\Eq(opt-joint)$$
the joint representation  in $\cK_{(x,y)}(t)$ such that
$$\cD_{\til{\r}}\left(
q^{\cT_1}[S^{(1)}_{(x,y)}|S^{(1)}_{V'\setminus\{(x,y)\}}(t-1)],
q^{\cT_2}[S^{(2)}_{(x,y)}|S^{(2)}_{V'\setminus\{(x,y)\}}(t-1)]\right) =$$
$$=\sum_{S^{(1)}_{(x,y)},S^{(2)}_{(x,y)}=1}^{8}
\til{\r} (S^{(1)}_{(x,y)},S^{(2)}_{(x,y)})\cdot
q^{*\; \cT_1,\cT_2}[S^{(1)}_{(x,y)},S^{(2)}_{(x,y)}
|S^{(1)}_{V'\setminus\{(x,y)\}}(t-1)),S^{(2)}_{V'\setminus\{(x,y)\}}(t-1)]
\; ;\Eq(q-star-locale)$$
the joint representation \equ(opt-joint)
can be exactly calculated as it is shown in Appendix A. All  probability 
measures introduced above depend on $\b,V',\s'$ and $\t'$; we drop explicit 
dependence to simplify the notation.
Notice that, of course, the $q^{\cT_i}[\cdot | \cdot]$ are stationary. The 
dependence on $t-1$ in the conditioning spins is made explicit only to clarify 
the updating rule from the configuration at time $t-1$ to the one at time $t$.
\smallno
5. Given $t\ge 1$ and the corresponding site $(x,y)\in V'$, the configurations 
$S^{(1)}(t)$ and $S^{(2)}(t)$ are obtained by choosing the pair
$(S^{(1)}_{(x,y)}(t),S^{(2)}_{(x,y)}(t))$ at random according to the joint 
probability measure \equ(opt-joint) and 
$S^{(i)}_{(x',y')}(t)=S^{(i)}_{(x',y')}(t-1)$ $\forall i=1,2$ and
$\forall (x',y')\in V'\setminus\{(x,y)\}$.
\medno
\par
In analogy with the terminology used in [DS1], we call this dynamics 
{\it Dynamical Surgery}; now some easy properties of this dynamics are 
discussed.
\medno
1. Given $t\in\bN$, it is immediate to see that if
$$
S^{(1)}_{(x',y')}(t-1)=S^{(2)}_{(x',y')}(t-1)\;\;\forall (x',y')\;
{\rm nearest\; neighbors\; of\;} (x,y)$$
then
$$
S^{(1)}_{(x,y)}(t)=S^{(2)}_{(x,y)}(t)\;\; ;
\Eq(prop-alg)
$$
in other words this means that in this case the joint representation
\equ(opt-joint) lies on the diagonal.
\smallno
2. The evolution of the two marginal chains
$S^{(1)}(t)$ and $S^{(2)}(t)$ is described by a standard Heat Bath 
dynamics; this is a consequence of the fact that the probability 
\equ(opt-joint) is a joint representation of the single site probability 
distributions \equ(marginali), which are the Gibbs measures 
of a single block--variable with boundary conditions given, respectively, 
by $\left(\cT_1,S^{(1)}_{V'\setminus\{(x,y)\}}(t-1)\right)$ and
$\left(\cT_2,S^{(2)}_{V'\setminus\{(x,y)\}}(t-1)\right)$.
Hence the equilibrium distributions for the two marginal chains
$S^{(1)}(t)$ and $S^{(2)}(t)$ are  the Gibbs measures with boundary 
conditions given, respectively, by $\cT_1$ and $\cT_2$.
\smallno
3. Given $t\in\bN$, given
$Q(\cdot,\cdot)\in\cK^{[\cT_1,\cT_2]_{(\bar{x},\bar{y})}}_{\b,\L',\s',\t'}$,
if the stochastic variable $X_t=(S^{(1)}(t),S^{(2)}(t))$ is such that
$$P(S^{(1)}(t)=S,S^{(2)}(t)=S')=Q(S,S')\;\;\forall S,S'\in\til{\O}_{V'}
\;\; ,$$
that is it is distributed according to the joint probability measure 
$Q(\cdot,\cdot)$, then it is easy to prove that $X_{t+1}$ is distributed 
according to an element of 
$\cK^{[\cT_1,\cT_2]_{(\bar{x},\bar{y})}}_{\b,\L',\s',\t'}$, as well.
\medno
\par
From the fact that the joint chain $X_t$ is an ergodic, aperiodic chain with a
finite  state space and from properties 2 and 3, it immediately follows that 
there exists a unique equilibrium  distribution for the Markov chain $X_t$; it 
is a joint probability measure $\bar{Q}(\cdot,\cdot)\in
\cK^{[\cT_1,\cT_2]_{(\bar{x},\bar{y})}}_{\b,\L',\s',\t'}$. This means that, 
provided the two marginal chains have thermalized, that is they are 
almost described by the corresponding Gibbs measures, the joint chain, after a 
long enough time, is almost characterized by the distribution 
$\bar{Q}(\cdot,\cdot)$,
which is a joint representation of the Gibbs measures 
$\m^{\cT_1}_{\b,\L',\s',\t'}(S)$ and $\m^{\cT_2}_{\b,\L',\s',\t'}(S)$.
Hence, if one needs to calculate the ``phase average" of an observable
depending on $S^{(1)}$ and $S^{(2)}$ with the joint probability $\bar{Q}$, 
one can perform a sufficiently long Monte Carlo run with the  dynamical 
surgery and calculate the ``time average" of the same observable. 
\par
If it were $\bar{Q}=Q^*$ then we could perform a Monte Carlo calculation of 
the uniqueness parameter 
$\cE^{[\cT_1,\cT_2]_{(\bar{x},\bar{y})}}_{\b,V',\s',\t'}$, but
in general $\bar{Q}\not= Q^*$ and moreover we cannot say anything on how much 
$\bar{Q}$ is close to $Q^*$. Anyway we can use
the joint representation $\bar{Q}$ to calculate an ``upper" estimator of 
the uniqueness parameter ;
we set:
$$\cU^{[\cT_1,\cT_2]_{(\bar{x},\bar{y})}}_{\b,V',\s',\t'}:={4\over l}
\cdot {1\over I}
\sum_{s=1}^{I} 
\r_{V'}^{}\left( S^{(1)}(s\cdot |V'|),S^{(2)}(s\cdot |V'|)\right)
\;\; \,\Eq(upper-estim)$$
where $I$ is the number of sweep performed during the Monte Carlo run.
It is clear that 
$\cU^{[\cT_1,\cT_2]_{(\bar{x},\bar{y})}}_{\b,V',\s',\t'}$ is a numerical
upper bound to $\cE^{[\cT_1,\cT_2]_{(\bar{x},\bar{y})}}_{\b,V',\s',\t'}$.
\par
The same Monte Carlo run can be used to estimate a ``lower" bound to the 
uniqueness parameter like in [BMO]. Indeed, we define the quantities
$$N^{(i)}_{(x,y)}(\a):={1\over I}
\sum_{s=1}^{I}\d_{\a,S^{(i)}_{(x,y)}(s\cdot |V'|)}
\;\;\forall i=1,2\;\forall (x,y)\in V'\;\forall\a\in\{1,...,8\}
\;\; \Eq(counter)$$ 
and we set
$$\cL^{[\cT_1,\cT_2]_{(\bar{x},\bar{y})}}_{\b,V',\s',\t'}:={4\over l}
\sum_{(x,y)\in V'} {1\over 2} \sum_{\a\in\{1,...,8\}}
|N^{(1)}_{(x,y)}(\a)-N^{(2)}_{(x,y)}(\a)|\;\; ;\Eq(low=estim)$$
see [BMO] for more details.
\bigno

%Sezione 5
\vfill\eject
\expandafter\ifx\csname sezioniseparate\endcsname\relax%
\input macro \fi
\numsec=5
\numfor=1\numtheo=1\pgn=1
\noindent
{\bf 5. Upper and lower bound: numerical results.}
\smallno
In Section 4 we have developed a Monte Carlo algorithm in order to evaluate
an  upper and a lower bound on $\cE^{[\cT_1,\cT_2]_{(x,y)}}_{\b,V',\s',\t'}$ 
and we have denoted these two quantities respectively by
$\cU^{[\cT_1,\cT_2]_{(x,y)}}_{\b,V',\s',\t'}$ and
$\cL^{[\cT_1,\cT_2]_{(x,y)}}_{\b,V',\s',\t'}$. 
\par
This algorithm cannot be 
used to give a direct evaluation of an upper bound to  $\cE_{\b,V'}$; indeed,
in order to calculate an upper bound to $\cE_{\b,V'}$ we should consider all 
possible constrained models and all possible pairs of boundary 
conditions $[\cT_1,\cT_2]_{(x,y)}$ (see the notation introduced in Section 2), 
that is we should perform $3\cdot l\cdot 2^{l^2+8l+1}$ runs of our Monte Carlo; 
one immediately realizes that this is an almost impossible task since it would
give rise to an enormous computation even in the case of small volumes  
$(l=2,3)$.
\par
Indeed, two boundary conditions can differ in one site 
$(x,y)\in\boundpiu{V'}$ in six ways: if one 
considers the block $B_{(x,y)}$ with $(x,y)\in\boundpiu{V'}$ and gives the 
block--variable $S_{(x,y)}$, only two of the original spins 
$\s^i_{(x,y)}$, those adjacent from the exterior to the lattice $V$, 
influence the equilibrium properties of the system; hence the possible 
ways in which two boundaries may differ are the following ones
$$\matrix{
+ + & + -\cr
+ + & - +\cr
+ + & - -\cr
+ - & - +\cr
+ - & - -\cr
- + & - -\cr
}\Eq(six-ways)$$
where we have depicted the two couples of original spins, belonging to the 
block $B_{(x,y)}$ and adjacent from the exterior to $V$, that one can have 
when one considers two 
boundary conditions differing just in $(x,y)$. Hence, one can easily convince 
himself that the total number of pairs of boundary conditions 
$[\cT_1,\cT_2]_{(x,y)}$ is given by $6\cdot 2^{2(4l-1)}$; finally, noticed that 
given $l$ there exists $2^{l^2}$ possible constrained models and that the site 
$(x,y)$ can be chosen in $4l$ different ways, it is immediate to see 
that considering all  possible constrained models and all possible 
pairs of boundary conditions amounts to examine $3\cdot l\cdot 2^{l^2+8l+1}$ 
different situations.
This number could be reduced by taking into account various symmetries;
however one would still have, for interesting values of $l$, an excessively big
computation to perform. \par 
Hence we are forced to consider just some of the
possible constrained models  and some of the possible boundary conditions, 
that is we are forced to perform  a sort of ``statistics".
\par
Let us denote by $\cU_{\b,V'}$ the highest value obtained  for 
$\cU^{[\cT_1,\cT_2]_{(x,y)}}_{\b,V',\s',\t'}$ and by
$\cL_{\b,V'}$ the corresponding estimate of
$\cL^{[\cT_1,\cT_2]_{(x,y)}}_{\b,V',\s',\t'}$. 
\par
First of all we  made a preliminary evaluation of our upper and lower 
estimators for different  values of $l$, by choosing completely at random a 
certain number of constrained models and some of all possible pairs of boundary 
conditions.
\par
We describe, now, how the statistics has been performed in all cases; our 
numerical results are summarized in Table 1.
\medno
$\bullet\; l=1$: In this case the volume $V'$ contains a single site, that is 
$V'=\{(1,1)\}$.
The dynamics defined in Section 4 is based on the local 
updating of a single block--variable; this updating is worked out according to 
the probability distribution \equ(opt-joint). Hence, in the case $l=1$ the 
estimator \equ(estim-mod-indep) has been exactly calculated
as follows:
$$\cE_{\b,V'}=4\Biggl(\sup_{\s'\in\O'_{V'}}\;
\sup_{(x,y)\in\boundpiu{V'}}\;\sup_{[\cT_1,\cT_2]_{(x,y)}}\;
\sum_{S^{(1)}_{(1,1)},S^{(2)}_{(1,1)}\in\{1,...,8\}\atop 
S^{(1)}_{(1,1)}\not= S^{(2)}_{(1,1)}}
q^{*\; \cT_1,\cT_2}[S^{(1)}_{(1,1)},S^{(2)}_{(1,1)}
]\Biggr)
;$$ 
On the other hand since in this case the Vasserstein distance coincides with 
the total variation distance (see Appendix A) we could compute the same 
quantity by means of the expression \equ (variation). We got:
$$
\cE_{\b,V'}= 2.119
$$
this result means that in this case there exists at least one constrained model
(we remark that there are only two constrained models, respectively 
corresponding to $\s'_{(1,1)}=+1$ and $\s'_{(1,1)}=-1$) and one pair of 
boundary conditions $[\cT_1,\cT_2]_{(x,y)}$ such that the Dobrushin--Shlosman 
condition is not fulfilled. Hence, the volume $V'$ with $l=1$ is not ``large 
enough" for our purposes.
\par
In both cases $\s'_{(1,1)}=+1$ and $\s'_{(1,1)}=-1$ and for many 
pairs of boundary conditions $[\cT_1,\cT_2]_{(x,y)}$, we have evaluated the 
estimator \equ(estim-vero) by means of the Monte Carlo algorithm, as well;
in this way we checked  
our computational procedure to get the best joint representation 
\equ (opt-joint).
The  results that we obtained differ from the exact values by 1--2\%; this 
shows that, at  least in the case $l=1$, our Monte Carlo procedure is very
efficient.
\medno
$\bullet\; l=2$: we have considered all  possible constrained models and for 
each model we have considered $200$ different pairs of boundary conditions. By 
performing $1.3\cdot 10^6$ full sweeps of our Monte Carlo, we have obtained the 
results in Table 1; these results show that there exists at least one 
constrained model $\s',\t'$ and one pair of boundary conditions 
$[\cT_1,\cT_2]_{(x,y)}$ such 
that $\cL^{[\cT_1,\cT_2]_{(x,y)}}_{\b,V',\s',\t'}>1$. This means that there
exists a  constrained model, the one corresponding to $\s',\t'$, which 
does not fulfill the Dobrushin--Shlosman uniqueness condition 
$DSU(V',\d)$ with $\d<1$.
\medno
$\bullet\; l=3$: we have considered all  possible constrained models and for 
each model we have considered $100$ different pairs of boundary conditions. The 
results in Table 1 refer to a run of $1.3\cdot 10^6$ sweeps. In this case 
there exists a particular constrained model $\s',\t'$ and a pair of boundary 
conditions $[\cT_1,\cT_2]_{(x,y)}$ such 
that $\cU^{[\cT_1,\cT_2]_{(x,y)}}_{\b,V',\s',\t'}>1$
and $\cL^{[\cT_1,\cT_2]_{(x,y)}}_{\b,V',\s',\t'}<1$; that is, the upper 
bound is ``too large", while the lower bound is ``too low". Hence, for this 
model we can neither say that the Dobrushin--Shlosman condition is fulfilled 
nor the opposite; we must necessarily consider larger volumes.
\medno
$\bullet\; l=4$: we have considered $50$ constrained model and $60$ pairs of 
boundary conditions; the results in Table 1 have been obtained by performing
$1.3\cdot 10^6$ full sweeps of the algorithm described in Section 4.
\medno
$\bullet\; l=8$: we have performed the same statistics as in the case $l=4$, 
but in this case it is obviously less significant, because the global number 
of possible choices is much greater. We have performed $1.3\cdot 10^6$ full 
sweeps of our Monte Carlo.
\medno
$\bullet\; l=16$: we have considered $30$ constrained model and $30$ pairs of 
boundary conditions; we had to reduce the number of runs, because of their 
length.
\medno
\par
\bigno
{\bf Remark.}
\par\noindent
The error $\D\cU_{\b,V'}$ in Table 1 is the usual statistical error, that is 
the standard deviation on the measure of the average $\cU_{\b,V'}$ of the 
Monte Carlo results. On the other hand the best estimate $\cL_{\b,V'}$ and the
error $\D\cL_{\b,V'}$ have been obtained by fitting the Monte Carlo results 
with the equation
$$\cL_{\b,V'} +{\bar \D\cL_{\b,V'}\over\sqrt I}\;\; ,$$
where $I$ is the number of full sweeps of the run (see [BMO])
and $\D\cL_{\b,V'} = {\bar \D\cL_{\b,V'}\over\sqrt I}$.
\bigno
\par
The results in Table 1 suggest that in the case $l=4$ all 
constrained models satisfy the Dobrushin--Shlosman condition, that is the 
volume $V'$ with $l=4$ is ``large enough" for our purposes.
But, strictly speaking, we cannot be sure about that, because we had to 
perform a statistics on the  constrained models and on the boundary conditions;
that is, there could exist a  particular ``bad" constrained model not 
satisfying the Dobrushin--Shlosman condition. Then we have considered values of
$l$ larger than $4$  and we have shown (see Table 1) that for $l=8,16$ the 
value of $\cU_{\b,V'}$ is so small 
and the effect of the change of conditioning spin is so localized 
(as we will explain) to lead one to the conclusion
that the existence of such a bad model can reasonably be excluded. 
\par
Our first observation refers to how  the quantity ${l\over 4} \cU_{\b,V'}$, 
namely the average  distance of the two copies of the system which evolve 
following the joint Monte  Carlo dynamics, behaves as a function of $l$. 
Indeed, it grows from $0.530$ to  $0.8415$ when one increases $l$ from $1$ to 
$3$, and then it remains approximately constant when one  increases the value 
of $l$; this is what one expects heuristically.
\par
In all the above described cases the ``statistics" has been performed by 
choosing completely at random the constrained models and the boundary 
conditions; in all cases we have found that the most ``dangerous" models, 
that is the constrained models with highest values of our upper estimator, are 
the striped and the chessboard ones.
\par
In order to strengthen the claim that  inequality \equ(estim-mod-indep) is 
satisfied provided one chooses $l$ large enough, in the case $l=6$ we have 
performed a ``rational" statistics, that is we have chosen the constrained 
models and the boundary conditions following reasonable criteria.
We have chosen $l=6$ 
for our final calculation, because the results listed in Table 1 suggest 
that in this case  condition \equ(estim-mod-indep) should be fulfilled, 
while, on the other hand, a full sweep of the Monte Carlo algorithm  takes
an acceptable CPU time so that we can perform a reasonably wide statistics.
\par
The algorithm introduced in Section 4 describes the ``coupled evolution" of
two  copies of the same model, characterized by two different boundary 
conditions; we recall that at each instant of time $t$ one and only one site
$(x,y)\in V'$ is updated and we observe that the property \equ(prop-alg) 
suggests that the differences between the two copies of the model have a
unique origin: the difference of the two boundary conditions in 
$(\bar{x},\bar{y})\in\boundpiu{V'}$.
\par
Due to this fact it seems resonable to assume that during the evolution the 
total number $\r_{V'}^{}(t):=\sum_{(x,y)\in V'} \left(1-
\d_{S^{(1)}_{(x,y)}(t),S^{(2)}_{(x,y)}(t)}\right)$
of disagreements between the two copies of the system is almost always equal to 
zero and sometimes these disagreements propagate in $V'$ starting from the site 
$(\bar{x},\bar{y})$ in $\boundpiu{V'}$. Now, if one recalls that the average 
distance between the two copies is approximately equal to $0.87$ for $l\ge 4$, 
it looks likely that the disagreements between the two copies of the system are 
localized in a ``small" region around $(\bar{x},\bar{y})$.
\par
We have tested this hypothesis as indicated in the histogram in Fig. 4
which shows the spatial dependence of disagreements between the two copies. We 
have plotted the histogram for various constrained models, moving the site 
$(\bar{x},\bar{y})$ along one of the four sides of $V'$ and for many pairs of 
boundary conditions; in all cases that we have considered, we have obtained 
histograms similar to that depicted in Fig. 4. The results summarized in 
Fig. 4 strongly suggest that the disagreements between the two copies of the 
model are almost completely 
localized in a suitably chosen $3\times 2$ rectangular block 
$R_{(\bar{x},\bar{y})}\subset V'$.
\par
Now, given the constrained model corresponding to $\s'\in\O'_{V'}$ and
$\t'\in\O'_{\boundpiu{V'}}$, the 
above remarks suggest that the average number of disagreements between 
the two copies of the system strongly depends on the values of 
$\s'_{(x,y)}$ with $(x,y)\in R_{(\bar{x},\bar{y})}$ and weakly on the values
of $\s'_{(x,y)}$ outside $R_{(\bar{x},\bar{y})}$. Hence, we have performed the 
statistics on the constrained model in the case $l=6$ by considering all
possible constrained models only inside $R_{(\bar{x},\bar{y})}$. In the 
following we precisely describe how the statistics has been performed.
\medno
$\bullet$ We have considered $(\bar{x},\bar{y})=(0,3)$,
$R_{(\bar{x},\bar{y})}=
\{(x,y)\in V' :\; x=1,2 \; {\rm and}\; 2\le y\le 4\}$, we have modified the 
boundary condition in $(\bar{x},\bar{y})$ in the six ways depicted in 
\equ(six-ways) and in each case we have considered two possible boundary 
conditions in 
$\boundpiu{V'}\setminus\{(\bar{x},\bar{y})\}$.
\smallno
$\bullet$ All possible constrained models have been considered in 
$R_{(\bar{x},\bar{y})}$, while in $V'\setminus R_{(\bar{x},\bar{y})}$
we have considered only the chessboard model and the model with 
$\s'_{(x,y)}=+1\;\forall (x,y)\in V'\setminus R_{(\bar{x},\bar{y})}$. Indeed, 
we expect that these two models are respectively the most and the less 
``dangerous" ones, as the results of the previous statistics suggest.
\smallno
$\bullet$ In each run of the joint Monte Carlo algorithm we have performed 
$10^5$ full sweeps, that is we have updated the whole lattice $V'$ $10^5$ 
times.
\medno
The results can be summarized by saying that the most dangerous constrained 
models inside $R_{(\bar{x},\bar{y})}$ appear to be the following ones
$$
\matrix{
+&-&+\cr
+&-&+\cr}\;\;\;\;\;\;\;\;\;
\matrix{
-&+&-\cr
-&+&-\cr}\;\;\;\;\;\;\;\;\;
\matrix{
+&-&+\cr
-&+&-\cr}\;\;\;\;\;\;\;\;\;
\matrix{
-&+&-\cr
+&-&+\cr}\;\; ;
\Eq(dangerous-model)
$$
in particular, our (indeed quite small)  statistics on the boundary conditions 
suggests that the most dangerous model among those in \equ(dangerous-model) is 
the second one; in this case, taking the chessboard model in $V' \setminus
R_{(\bar{x},\bar{y})}$  we obtain
$\cU^{[\cT_1,\cT_2]_{(\bar{x},\bar{y})}}_{\b,V',\s',\t'}=0.610$. The numerical
results confirm the weak dependence of the estimators on the constrained model 
outside $R_{(\bar{x},\bar{y})}$, as well; actually the differences are of  
5-10\%. Finally, this set of Monte Carlo runs shows that the most dangerous  
ways in which one can modify the boundary conditions in $(\bar{x},\bar{y})$  
are the following ones
$$++\;\; - - \;\;\;\; {\rm and}\;\;\;\; + -\;\; - +\;\; .\Eq(dang-bound)$$ 
\par
Once we have understood what are the worst situations inside 
$R_{(\bar{x},\bar{y})}$ and in $(\bar{x},\bar{y})$ we have performed the wide 
statistics on the possible boundary conditions described below.
\medno
$\bullet$ We have considered the most dangerous constrained model inside
$R_{(\bar{x},\bar{y})}$.
\smallno
$\bullet$ We have considered ten possible constrained model outside
$R_{(\bar{x},\bar{y})}$; six of them are those depicted in Fig. 1, the 
remaining four have been chosen at random.
\smallno
$\bullet$ In each case we have considered twenty possible pairs of boundary 
conditions with the original spins in $B_{(\bar{x},\bar{y})}$ and adjacent to 
$V$ chosen like in \equ(dang-bound).
\medno
The weak dependence of the estimator on the constrained models outside 
$R_{(\bar{x},\bar{y})}$ has been confirmed and we have found
$$
\cU_{\b,V'}=0.633\;\;\; \D\cU_{\b,V'}=0.011\;\; ;\Eq(res-L6)$$
hence we can confidently say that the condition \equ(estim-mod-indep)
is satisfied if one considers the volume $V'$ with $l=6$.
\bigno

%Sezione 6
\vfill\eject
\expandafter\ifx\csname sezioniseparate\endcsname\relax%
\input macro \fi
\numsec=6
\numfor=1\numtheo=1\pgn=1
\noindent
{\bf 6. Conclusions.}
\smallno
As we explained in Section 2  the problem of proving 
Gibbsianness of our  renormalized measure is reduced to the verification of
$DSU(V,\d)$ condition for some $V$ and 
$\d <1$ for all possible constrained models.\par
It is clear that disproving the condition for a given volume $V$ is much easier
than proving it since, to disprove, it is sufficient to exhibit one constrained
model and one boundary condition for which a lower bound $\cL$ for the 
uniqueness parameter appearing in $DSU$ exceeds one; moreover since this lower 
bound involves variation distance and then thermal averages it is naturally 
computable via a Monte Carlo procedure. On the other hand an upper bound has 
to involve the consideration of all possible constrained models as well as all 
possible boundary conditions. Moreover a priori it was not clear how to 
provide an upper bound based on a Monte Carlo computation; this motivated the 
idea of the dynamical surgery. The necessity of a Monte Carlo approach comes 
from the consideration of how fast the number of constrained models and 
possible boundary conditions grows as a function of $l$, the side of the 
squared volume where we try to verify $DSU$.
\par
Since for $l=2$ we find a particular constrained model and boundary condition 
for which $\cL$ is greater than one, necessarily we have to go at least to 
$l=3$ and already the number of independent computations is very large.
Moreover since for $l=3$ the lower bound seems always less than one whereas 
for at least one case the upper bound is larger than one we can neither 
disprove nor try to prove with our bounds the validity of $DSU$ for a square 
with side 3 so that we have to go at least to $l=4$.
\par 
It appears clear from our numerical computations that our upper estimator
$\cU$ for the uniqueness parameter has the correct behaviour with $l$: as a
consequence of the spatial localization of the set of disagreements between 
our two coupled processes  $\cU$ is inversely proportional to $l$; thus 
increasing $l$ is the correct choice to get a value of  $\cU$ so smaller than 
one (included the error) to be confident in the validity of $DSU$. 
Unfortunately increasing $l$ implies an enormous increase in the number of 
computations; introducing some statistics becomes necessary.
The right compromise between smallness of $\cU$ and number of constrained 
models and boundary conditions came out to be $l=6$.
In this case we performed the ``rational statistics" that we have described in
Section 5 by exploiting the (numerically evident) small dependence of 
$\cU$ on the constrained model and on the boundary condition far from 
conditioning spin that we are changing.
\par 
Our computations relative to the case $l=6$ make us really confident on the
uniform validity of $DSU$.
\par
We can say that our method is successful because even the
correlation length of the ``worst" constrained models is very small; however 
it is not so small  to avoid the consideration of sides $l$ at least greater 
than 4.
\par
We want to make now some general remarks on our Monte Carlo algorithm.
\par
As it has been underlined in [DS1], the nice feature of finite size conditions 
like $DSU$, involving the behavior of Gibbs measures in finite volumes,
is that they are able to imply  absence of phase
transitions and many nice properties of the unique {\it infinite volume} Gibbs
state. This point of view is very helpful for example when we have to decide 
whether or not a given system is in the unique phase regime; especially when 
we do not have a natural parameter (like the inverse temperature $\b$) whose 
smallness imply weak-coupling.
\par
The ``uniqueness test", based on the verification of $DSU$, has the advantage,
w.r.t the traditional Monte Carlo test, of being based on rigorous grounds. 
However a real computer assisted proof seems very difficult to achieve unless 
the concerned volumes are really very small. If this is not the case the use 
of a Monte Carlo algorithm becomes essential; then the situation is somehow 
intermediate between a traditional Monte Carlo simulation and a computer 
assisted proof.
\par
We want to remark that our algorithm to compute a numerical upper bound on the
Vasserstein distance between Gibbs measures, which is based on 
``local readjustment" of the coupling, seems to be quite performant and 
probably it can be used in more general contexts. 
Finally it is remarkable that, due to the very nature of the
coupling procedure, the statistical error on $\cU$ is much smaller that the
corresponding one on $\cL$; indeed in this last case {\it all} sites of our 
volume and not only the disagreements, like in the computation of $\cU$, play 
a role as a source of statistical errors.
\bigno

%Acknowledgements.
\bigno
\bigno
\bigno
\bigno
\noindent
{\bf Acknowledgements.}
\smallno
We are indebted to Leonardo Angelini, Giosi Benfatto, Paolo Cea, 
Leonardo Cosmai, Tom Kennedy, Mario Pellicoro and especially to Aernout van
Enter for very stimulating and illuminating discussions.
We also thank Christian Maes and Enzo Marinari for helpful conversations.
We want to express thanks to Istituto Nazionale di Fisica Nucleare - Sezione 
di Bari whose financial support made possible this collaboration. This work 
has been partially supported by the grant CHRX-CT93-0411 and CIPA-CT92-4016 of 
the Commission at European Communities.
\bigno

%Appendice A
\vfill\eject
\expandafter\ifx\csname sezioniseparate\endcsname\relax%
\input macro \fi
\numsec=7
\numfor=1\numtheo=1\pgn=1
\noindent
{\bf 7. Appendix A.}
\smallno
We consider the space $S:=\{1,...,n\}$ with $n\ge 2$ and the metrics
$\r (s,s')\equiv\til{\r}(s,s'):=1-\d_{s,s'}$ $\forall s,s'\in S$; let us 
denote by $\l$ and $\m$ two probability measures on $S$ and by $\cK$ the 
set of the joint representations of $\l$ and $\m$. Hence, given $q\in\cK$ one 
has that $q$ is a probability measure on $S\times S$ such that
$$\sum_{s\in S} q(s,s')=\m(s')\;\forall s'\in S
\;\; {\rm and}\;\; 
\sum_{s'\in S} q(s,s')=\l(s)\;\forall s\in S \;\; .
\eqno(A.1)$$
\par
We recall, now, that the {\it total variation distance} and the 
{\it Vasserstein distance} between $\l$ and $\m$ are respectively given by
$$\eqalign{
{ Var} (\l,\m)&:={1\over 2}\sum_{s\in S}|\l(s)-\m(s)|\cr
\cD_{\r}(\l,\m)&:=\inf_{q\in\cK}\sum_{s,s'\in S} \r(s,s')\cdot
q(s,s')\cr}\eqno(A.2)$$
\par
\medno
{\bf Proposition A1.}
\smallno
1. With the notation introduced before
$$\cD_{\r}(\l,\m)={ Var}(\l,\m)\eqno(A3)$$
\smallno
2. Let us consider the following partition of the set $S$
$$S=A\cup B\cup C\;\; ,\eqno(A.4)$$
where
$$\eqalign{
A&:=\{s\in S:\; \l(s)>\m(s)\}\cr
B&:=\{s\in S:\; \l(s)<\m(s)\}\cr
C&:=\{s\in S:\; \l(s)=\m(s)\}\cr}
\eqno(A.5)$$
one has that
$${  Var}(\l,\m)=\sum_{s\in A} (\l(s)-\m(s))=
\sum_{s\in B} (\m(s)-\l(s))\;\; .\eqno(A.6)$$
\medno
{\it Proof.}
\smallno
1. See [D1], page 472.
\smallno
2. It is an immediate consequence of the normalization of $\m$ and $\l$.
$\square$
\par
We want, now, to calculate the particular joint representation $q^*\in\cK$ such 
that the following equality is satisfied
$$\cD_{\r}(\l,\m)=\sum_{s,s'\in S} \r(s,s')\cdot q^*(s,s')\;\; ;
\eqno (A.7)$$
that is we are looking for the joint representation ``optimizing" the sum in 
the definition of $\cD_{\r}(\l,\m)$. 
In other words, by virtue of the above Proposition, we can say that
our aim is to find an $n\times n$ square matrix 
$$q^*=\left(
\matrix{
q^*_{1,n}&q^*_{2,n}&q^*_{3,n}&...&q^*_{n,n}\cr
...      &...&...&...&...      \cr
...      &...&...&...&...      \cr
q^*_{1,3}&q^*_{2,3}&q^*_{3,3}&...&q^*_{n,3}\cr
q^*_{1,2}&q^*_{2,2}&q^*_{3,2}&...&q^*_{n,2}\cr
q^*_{1,1}&q^*_{2,1}&q^*_{3,1}&...&q^*_{n,1}\cr}
\right)\;\; ,\eqno(A.8)$$
such that 
$$\sum_{s=1}^{n} q^*_{s,s'}=\m_{s'}\;\forall s'=1,...,n\;\;\;
{\rm and}\;\;\;
\sum_{s'=1}^{n} q^*_{s,s'}=\l_s\;\forall s=1,...,n\;\;
\eqno(A.9)$$
and such that the sum of the off--diagonal elements is given by
$$\sum_{s,s':\; s\not= s'} q^*_{s,s'}=
\sum_{s\in A} (\l_s-\m_s)=
\sum_{s\in B} (\m_s-\l_s)\;\;\; ;\eqno(A.10)$$
we have introduced the notation: $\l_s:=\l(s)$ and $\m_s:=\m(s)$ 
$\forall s\in S$.
\par
In order to compute all the elements of the matrix $q^*_{s,s'}$ one can 
follow the procedure described below; first of all we introduce the following
{\it filling rule}. It is a construction of a matrix $q^*$ along a sequence of 
steps. At each step some new set of entries of $q^*$ will be determined whereas
for the rest of the entries only some suitable sums of them will be determined.
We will end up with a matrix $q^*$ satisfying both $(A.9)$, $(A.10)$ and in 
this way we show that our procedure is well defined in the sense that at each 
step it is not empty the set of matrices satisfying our requirements.
\medno
1. If $\l_s=\m_s\;\forall s\in S$ then 
$$q^*_{s,s'}=\l_s\d_{s,s'}\;\;\;\forall s,s'\in S $$
\smallno
2. If $\exists s\in S$ such that $\l_s\not=\m_s$ then $\forall s\in S$
one puts
$$q^*_{s,s}=\min\{\l_s,\m_s\}$$
and
$$\eqalign{
{\rm if}\;\m_s=\l_s\; {\rm then}\; q^*_{s,s'}=0&\;\forall s'\not= s,
                                   \; q^*_{s',s}=0\;\forall s'\not= s\cr
{\rm if}\;\m_s>\l_s\; {\rm then}\; q^*_{s,s'}=0&\;\forall s'\not= s
	\; {\rm and}\; q^*_{s',s}\;{\rm are\; subject \; to \; the \; 
        condition}:\cr
                &\sum_{s':\; s'\not=s}q^*_{s',s}=\m_s-\l_s\cr 
{\rm if}\;\l_s>\m_s\; {\rm then}\; q^*_{s',s}=0&\;\forall s'\not= s
	\; {\rm and}\; q^*_{s',s}\;{\rm are\; subject \; to \; the \; 
        condition}:\cr
                &\sum_{s':\; s'\not=s}q^*_{s,s'}=\l_s-\m_s\cr}
$$
\medno
In Fig. 5 an example of application of the above filling rule is 
shown.
\par
In case 1 the problem is trivially solved; indeed by means of the above 
filling rule one obtains a diagonal matrix fulfilling the requirements $(A.9)$ 
and $(A.10)$. In case 2 it is immediate to see that we construct a matrix 
satisfying the requirements $(A.9)$, but the problem is that in general not 
all off--diagonal terms are determined: many of them are set equal to 
zero, but for the others only some sums are specified and not the single 
values.
\par
By collecting all  off--diagonal non--vanishing entries  
one obtains an $|A|\times |B|$ rectangular matrix, whose elements are not yet 
determined. 
More correctly we should speak about a class of matrices (which at the end of 
our construction will be shown to be non--empty):
the ones satisfying the requirements corresponding to one step of application 
of our filling rule.
\par
Let us denote by $q^{(1)}_{i,j}$ with $1\le i\le n_1:=|A|$ and  
$1\le j\le n'_1:=|B|$ the elements of this new matrix. We
observe that $q^{(1)}_{i,j}$ corresponds to the element $q^*_{f(i),f'(j)}$ of 
the original matrix, where
$$\eqalign{
f(1)&:=\min\{s\in S:\;\l_s>\m_s\}\cr
f(i)&:=\min\{s>f(i-1):\;\l_s>\m_s\}\;\;\;\forall i>1\cr}\;\; \eqno(A.11)$$
and
$$\eqalign{
f'(1)&:=\min\{s\in S:\;\m_s>\l_s\}\cr
f'(i)&:=\min\{s>f'(i-1):\;\m_s>\l_s\}\;\;\;\forall i>1\cr}\;\; .\eqno(A.12)$$
\par
It is important to observe that, due to the filling rule, the elements of the 
new matrix must satisfy the following requirements
$$\sum_{j=1}^{n'_1} q^{(1)}_{i,j}=\l_{f(i)}-\m_{f(i)}\;\forall i=1,...,n_1
\;\; {\rm and}\;\;
\sum_{i=1}^{n_1} q^{(1)}_{i,j}=\m_{f'(j)}-\l_{f'(j)}\;\forall i=1,...,n_1
\; .\eqno(A.13)$$
From the fact that $q^{(1)}$ has been constructed by collecting all 
off--diagonal elements of $q^*$ different from zero and from
equations $(A.13)$, it follows that the $(A.10)$ is fulfilled.
\par
In order to fill the matrix $q^{(1)}$ (by further reducing its 
indetermination) one applies again the filling  rule stated above. Again two 
things may happen: all elements of $q^{(1)}$ are fixed or one obtains another 
$n_2\times n'_2$ rectangular matrix $q^{(2)}$ to be filled. 
\par
One goes on applying the filling rule and in this way one 
obtains a sequence of rectangular matrices 
$$q^*,q^{(1)},q^{(2)},...,q^{(t)},...$$ 
which are smaller and smaller in the sense that 
$$n\cdot n > n_1\cdot n'_1 > n_2\cdot n'_2 > ...\;\;\; ;$$ 
the procedure is stopped when all  elements are fixed.
\bigno

%Bibliografia
\vfill\eject
\noindent
\centerline{\bf References.}
\smallno
\item{[BMO]}
G. Benfatto, E. Marinari and E. Olivieri, 
``Some numerical results on the block spin transformation for the 2D Ising 
model at the critical point", Journ. Stat. Phys. {\bf 78}, 731--757 (1995).
\item{[C]}
C. Cammarota, ``The Large Block Spin Interaction", Il Nuovo Cimento
{\bf 96B}, 1--16 (1986).
\item{[CG]}
M. Cassandro, G. Gallavotti, ``The Lavoisier Law and the
Critical Point", Il Nuovo Cimento {\bf 25B}, 691 (1975).
\item{[D1]} 
R. L. Dobrushin, ``Prescribing a system of random variables by
conditional distributions", Theory of Prob. Appl. {\bf 15}, 453--486
(1970).
\item{[D2]} 
R. L. Dobrushin, Lecture given at the workshop: ``Probability and Physics"
Renkum, Holland, (1995).
\item{[DS1]}
R. L. Dobrushin, S. Shlosman, ``Constructive Criterion for the
Uniqueness of Gibbs Fields", Stat. Phys. and Dyn. Syst., 
Birkhauser, 347--370 (1985).
\item{[DS2]}
R. L. Dobrushin, S. Shlosman, ``Completely Analytical Gibbs
Fields", Stat. Phys. and Dyn. Syst., Birkhauser, 371--403 (1985).
\item{[DS3]}
R. L. Dobrushin, S. Shlosman, ``Completely Analytical
Interactions Constructive Description", Journ. Stat.
Phys. {\bf 46}, 983--1014 (1987).
\item{[E1]}
A. C. D. van Enter, ``Ill--defined block--spin transformations at arbitrarily 
high temperatures", Journ. Stat. Phys., {\bf 83} to appear, (1996). 
\item{[E2]}
A. C. D. van Enter, ``On the possible failure of the Gibbs property for 
measures on lattice systems", preprint (1996).
 \item{[EFK]}
A. C. D. van Enter, R. Fern\'andez, R. Koteck\'y,
``Pathological behavior of renormalization group maps at high field and above 
the transition temperature", Journ. Stat. Phys., {\bf 79}, 969--992 (1995).
 \item{[EFS]}
A. C. D. van Enter, R. Fern\'andez, A. D. Sokal,
``Regularity Properties and Pathologies of Position--Space
Renormalization--Group Transformations: Scope and Limitations of
Gibbsian Theory", Journ. Stat. Phys. {\bf 72}, 879--1167 (1994).
\item{[FP]}
R. Fernandez, C. Pfister, ``Non Quasi--Local Projections of Gibbs
States", preprint (1994).
\item{[GP]}
R. B. Griffiths, P. A. Pearce, ``Mathematical Properties of Position--Space
Renormalization Group Transformations", Journ. Stat. Phys. {\bf 20}, 499--545 
(1979).
\item{[I]}
R. B. Israel, ``Banach Algebras and Kadanoff Transformations
in Random Fields", J. Fritz, J. L. Lebowitz and D. Szasz editors
(Esztergom 1979), Vol. II, 593--608 (North--Holland, Amsterdam 1981).
\item{[HK]}
K. Haller, T. Kennedy,
``Absence of renormalization group pathologies near the critical 
temperature--two examples", University of Arizona Preprint, Austin Archives 
95--505, (1995).
\item{[K1]}
T. Kennedy, ``A fixed point equation for the high temperature phase of discrete
lattice spin systems", Journ. Stat. Phys. {\bf 59}, 195 (1990).
\item{[K2]}
T. Kennedy, ``Some Rigorous Results on Majority Rule Renormalization Group
Transformations near the Critical Point", Journ. Stat. Phys. {\bf 72}, 15
(1993).
\item{[K3]}
T. Kennedy, private communication.
\item{[Ko]}
O. K. Kozlov, ``Gibbs Description of a System of Random Variables",
Probl. Inform. Transmission. {\bf 10}, 258--265 (1974).
\item{[LV]}
J. Lorinczi, K. Vande Velde, ``A Note on the Projection of Gibbs
Measures", Journ. Stat. Phys. {\bf 77}, 881--887 (1994).
\item{[Ma]} Ch. Maes, private communication.
\item{[MO1]}
F. Martinelli, E. Olivieri, ``Finite Volume Mixing Conditions for Lattice
Spin Systems and Exponential Approach to Equilibrium of Glauber
Dynamics", Proceedings of 1992 Les Houches Conference
on  Cellular Automata and Cooperative Systems, N. Boccara, E. Goles, S.
Martinez and P. Picco editors (Kluwer 1993).
\item{[MO2]}
F. Martinelli, E. Olivieri, ``Approach to Equilibrium of Glauber Dynamics
in the One Phase Region I. The Attractive Case", 
Commun. Math. Phys. {\bf 161}, 447--486 (1994).
\item{[MO3]}
F. Martinelli, E. Olivieri, ``Approach to Equilibrium of Glauber Dynamics
in the One Phase Region II. The General Case",
 Commun. Math. Phys. {\bf 161}, 487--514 (1994).
\item{[MO4]}
F. Martinelli, E. Olivieri, ``Some Remarks on Pathologies of
Renormalization Group Transformations for the Ising model",
Journ. Stat. Phys. {\bf 72}, 1169--1177 (1994).
\item{[MO5]}
F. Martinelli, E. Olivieri, ``Instability of renormalization group pathologies
under decimation", 
Journ. Stat. Phys. {\bf 79}, 25--42 (1995).
\item{[MOS]}
F. Martinelli, E. Olivieri, R. Schonmann, ``For $2$--$D$ Lattice Spin
Systems Weak Mixing Implies Strong Mixing",
Commun. Math. Phys. {\bf 165}, 33--47 (1994).
\item{[NL]} 
Th. Niemeijer, M. J. van Leeuwen, ``Renormalization theory
for Ising--like spin systems", in ``Phase Transitions and 
Critical Phenomena", vol. 6, Eds. C. Domb, M. S. Green 
(Academic Press, 1976).
\item{[O]}
E. Olivieri, ``On a Cluster Expansion for Lattice Spin Systems: a Finite
Size Condition for the Convergence",
Journ. Stat. Phys. {\bf 50}, 1179--1200 (1988).
\item{[OP]}
E. Olivieri, P. Picco, ``Cluster Expansion for $D$--Dimensional Lattice
Systems and Finite Volume Factorization Properties",
Journ. Stat. Phys. {\bf 59}, 221--256 (1990).
\item{[S]}
R. H. Schonmann, ``Projections of Gibbs Measures May Be Non--Gibbsian",
Commun. Math. Phys. {\bf 124}, 1--7 (1989).
\item{[Sh]}
S. Shlosman, ``Uniqueness and half--space non--uniqueness of Gibbs states in 
Czech models", Teor. Math. Phys. {\bf 66}, 284--293 (1986).

%Tabella 1
\vfill\eject
\centerline{\bf Table 1}
\vskip 2 truecm
\hfil{     % Questo hfil serve a centrare la tabella
\vbox{\offinterlineskip % Offinterlineskip annulla il lineskip
\hrule     %traccia una riga orizzontale
{\phantom .}
\hrule     %traccia una riga orizzontale
\halign{&\vrule#&
   \strut\quad\hfil#\quad\cr
height2pt
&\omit&&\omit&&\omit&&\omit&&\omit&\cr
&\hfil $l$\hfil&
&\hfil $\cU_{\b,V'}$\hfil&
&\hfil $\D\cU_{\b,V'}$\hfil&
&\hfil $\cL_{\b,V'}$\hfil&
&\hfil $\bar \D\cL_{\b,V'}$\hfil&\cr
height2pt&\omit&&\omit&&\omit&&\omit&&\omit&\cr
\noalign{\hrule}
\noalign{
{\phantom .}
\hrule}     %traccia una riga orizzontale
height2pt&\omit&&\omit&\cr
&2&& 1.452 && 0.0074 && 1.34 && -2.50&\cr
&3&& 1.122 && 0.0071 && 0.90 && +2.05&\cr
&4&& 0.877 && 0.0046 && 0.73 && +0.32&\cr
&8&& 0.436 && 0.0026 && 0.30 && +0.74&\cr
&16&&0.207 && 0.0016 && 0.16 && +0.04&\cr
height2pt&\omit&&\omit&\cr}
\hrule
{\phantom .}
\hrule     %traccia una riga orizzontale
}}
\vskip 5 truecm
\noindent
\centerline{
\vbox{\hsize=12 truecm\baselineskip=10 pt
\noindent
Table 1: In the table we have listed the values of
$\cU_{\b,V'}$, $\D\cU_{\b,V'}$, $\cL_{\b,V'}$ and $\bar\D\cL_{\b,V'}$ obtained 
at $l=2,3,4,8,16$. The meaning of  
$\D\cU_{\b,V'}$ and $\bar \D\cL_{\b,V'}$ is explained in the Remark in Section
5.
}}

%Figura 1
\vfill\eject
\midinsert
\vskip 20 truecm
\includegraphics{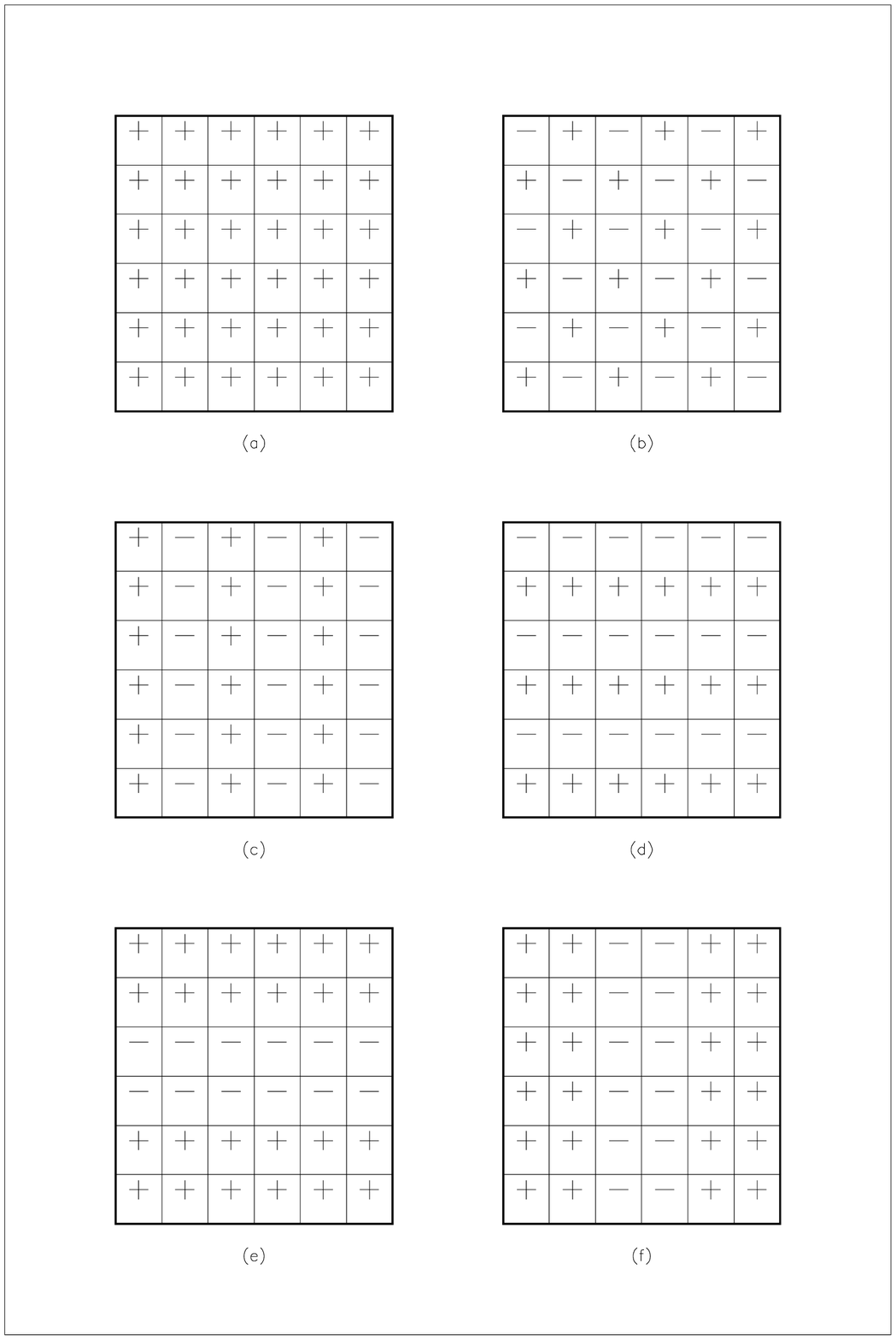}
\endinsert
\vskip 0.7 truecm
\centerline{
\vbox
{
\hsize=13truecm
\baselineskip 0.35cm
\noindent
{\ninebf Fig.1:}{\ninerm \quad
Examples of configurations of renormalized variables corresponding 
to constrained models which are particularly important in our calculations.}
}}

%Figura 2
\vfill\eject
\midinsert
\vskip 10 truecm
\includegraphics{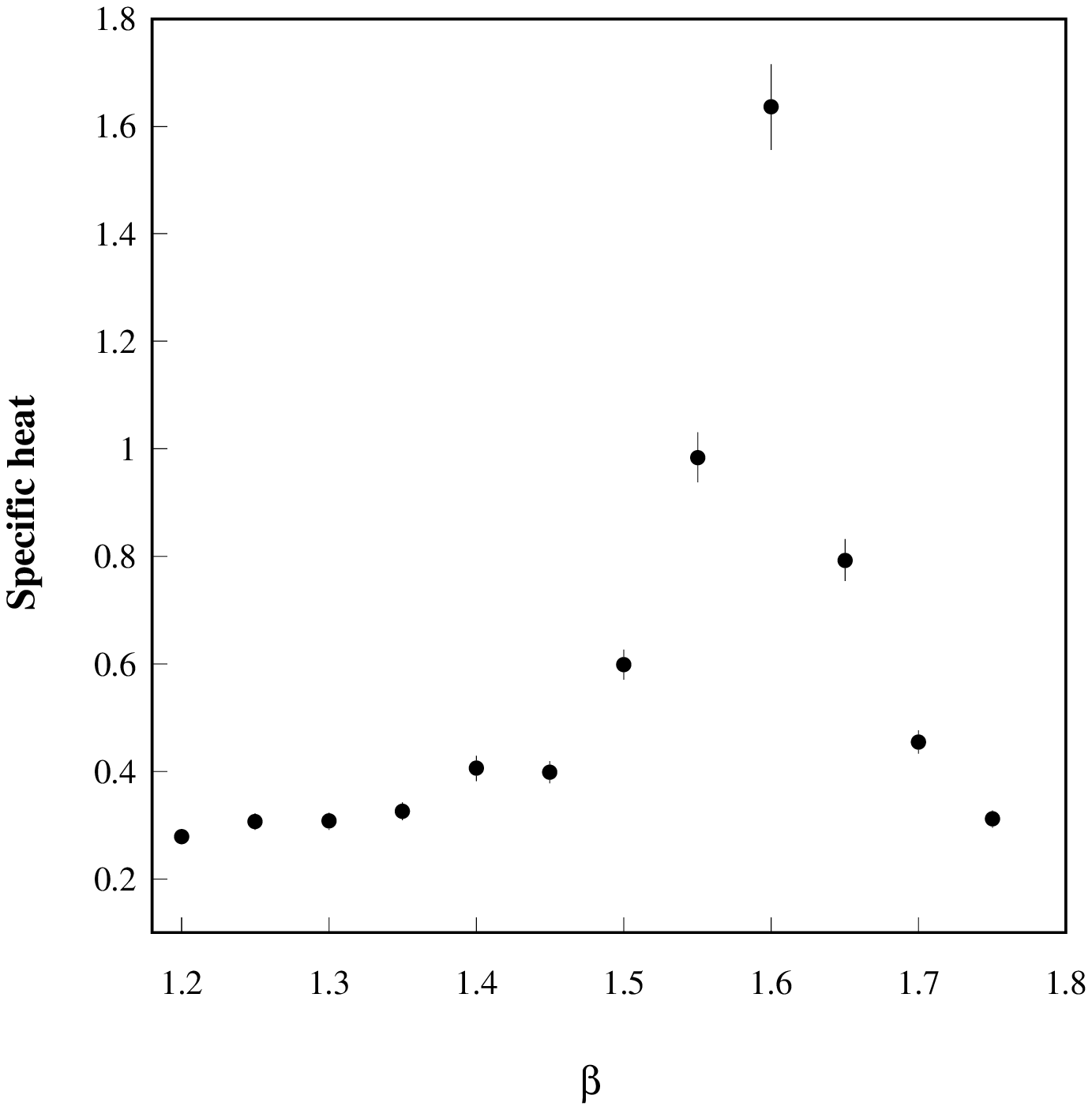}
\endinsert
\vskip 10 truecm
\centerline{
\vbox
{
\hsize=13truecm
\baselineskip 0.35cm
\noindent
{\ninebf Fig.2:}{\ninerm \quad
Specific heat for the chessboard model defined in \equ(chess-stri) 
as a function of the inverse temperature $\b$.}
}}

%Figura 3
\vfill\eject
\midinsert
\vskip 10 truecm
\includegraphics{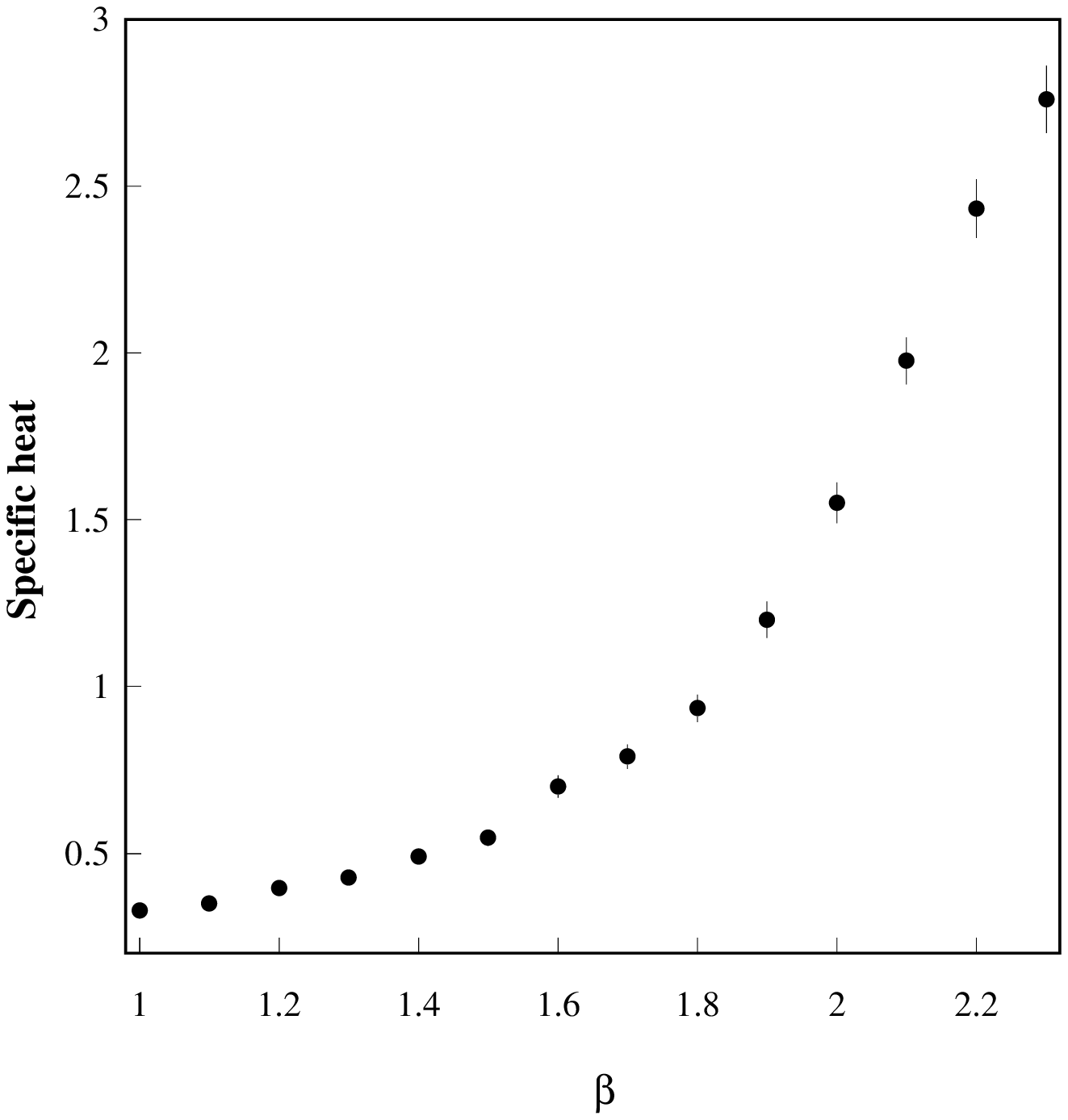}
\endinsert
\vskip 10 truecm
\centerline{
\vbox
{
\hsize=13truecm
\baselineskip 0.35cm
\noindent
{\ninebf Fig.3:}{\ninerm \quad
Specific heat for the striped model defined in \equ(chess-stri) 
as a function of the inverse temperature $\b$.}
}}

%Figura 4
\vfill\eject
\midinsert
\vskip 10 truecm
\includegraphics{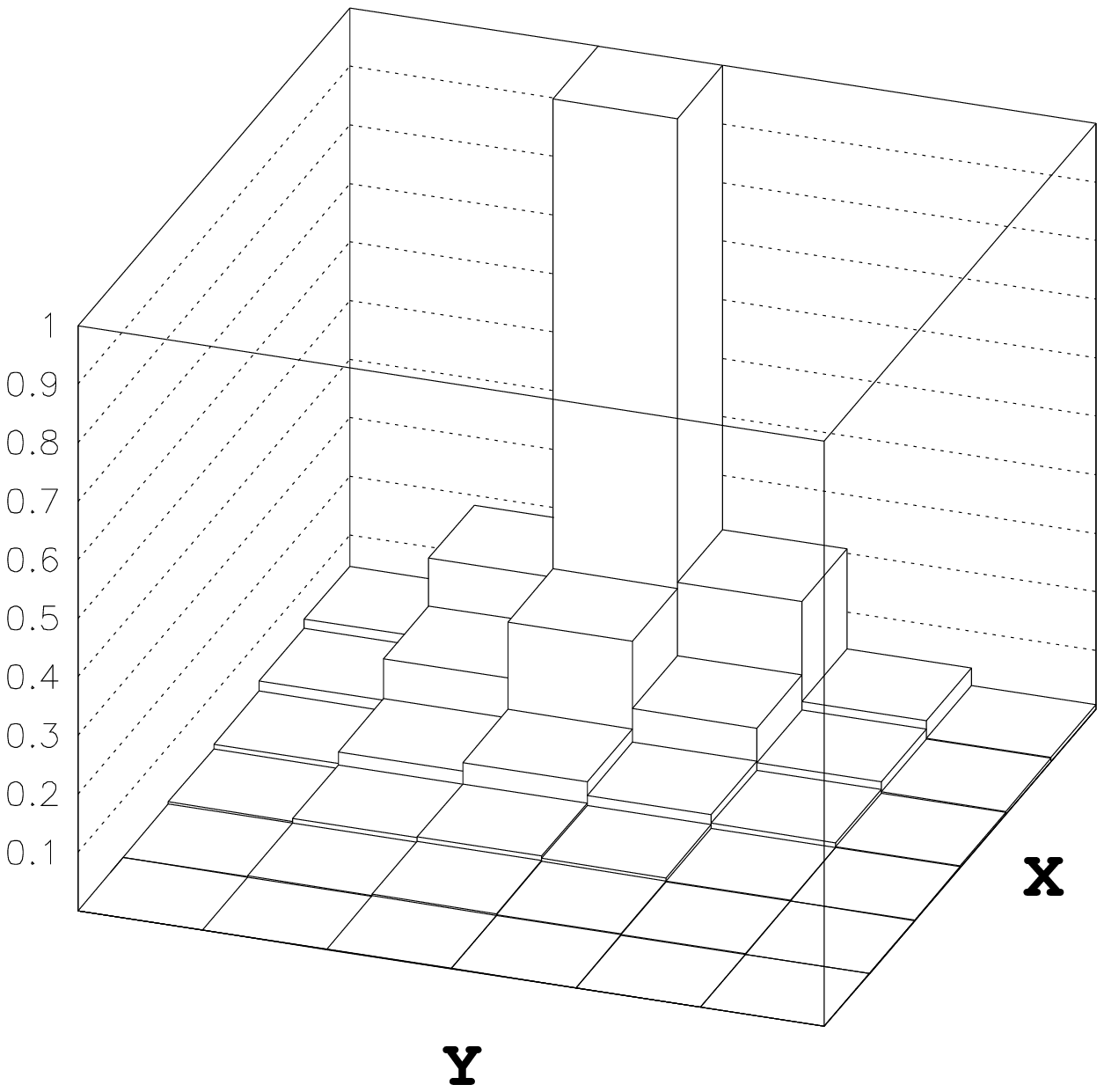}
\endinsert
\vskip 4 truecm
\centerline{
\vbox
{
%\hsize=16truecm
%\baselineskip 0.35cm
\noindent
{\ninebf Fig.4:}{\ninerm \quad
Each square of the plane {\bf X--Y} represents a site $(x,y)$ of a 
lattice $V'$ with $l=6$; the {\bf X} axis is oriented from the left to 
the right, while the {\bf Y} axis is oriented from the top of the figure to the 
bottom. The results contained in the histogram refer to the chessboard model: 
the height of each column $(x,y)$ is given by the ratio 
$${\sum_{s=1}^I \sum_{(x,y)\in V'}(1-\d_{S^{(1),s}_{(x,y)},S^{(2),s}_{(x,y)}})
\over
\sum_{s=1}^I \sum_{(x,y)\in V'}(1-\d_{S^{(1),s}_{(1,3)},S^{(2),s}_{(1,3)}})}$$
where $I=1\cdot 10^5$ is the number of full sweeps of the run, 
$S^{(1),s}_{(x,y)}$ and $S^{(2),s}_{(x,y)}$ are the block--variables of the 
two copies of the system on the site $(x,y)$ and after $s$ sweeps. The results 
in the histogram have been obtained considering a pair of boundary conditions 
$[\cT_1,\cT_2]_{(0,3)}$, such that the two pairs of original spins in the 
block $B_{(0,3)}$ are $+-$ and $-+$.}
}}

%Figura 5
\vfill\eject
$\phantom {.}$
\vskip 4 truecm
\par\noindent
$$\matrix{
\m_5& &   0   &   0   &   0   &   0   & \m_5  \cr
\m_4& &\square&   0   &\square& \l_4  &\square\cr
\m_3& &   0   &   0   & \m_3  &   0   &   0   \cr
\m_2& &\square& \l_2  &\square&   0   &\square\cr
\m_1& & \m_1  &   0   &   0   &   0   &   0   \cr
    & &       &       &       &       &       \cr
    & & \l_1  & \l_2  & \l_3  & \l_4  & \l_5  \cr
    & &       &       &       &       &       \cr
    & &       &       &       &       &       \cr
    & &       &       &  (a)  &       &       \cr}
\;\;\;\;\;\;\;\;\;\;\;\;\;\;\;\;\;
\matrix{
q^{(1)}_{1,2}&q^{(1)}_{2,2}&q^{(1)}_{3,2}&\m_4-\l_4\cr
q^{(1)}_{1,1}&q^{(1)}_{2,1}&q^{(1)}_{3,1}&\m_2-\l_2\cr
             &             &             &           \cr
  \l_1-\m_1  &  \l_3-\m_3  &  \l_5-\m_5  &           \cr
             &             &             &           \cr
             &             &             &           \cr
             &     (b)     &             &           \cr}
$$
\vskip 6 truecm
\centerline{
\vbox
{
%\hsize=16truecm
%\baselineskip 0.35cm
\noindent
{\ninebf Fig.5:}{\ninerm \quad
Application of the filling rule in the case: $n=5$, $\l_1>\m_1$, 
$\l_2<\m_2$, $\l_3>\m_3$, $\l_4<\m_4$ and $\l_5>\m_5$. In $(a)$ we have 
depicted the matrix $q^*_{s,s'}$: the sum of the elements of one of its rows 
must be equal to the corresponding value depicted on the left, while the 
sum of the elements of one of its column must be equal to the corresponding 
value depicted on the bottom. The squares represent elements which are not 
fixed after the filling rule is applied for the first time. The collection 
of all  squares gives rise to the matrix $q^{(1)}_{i,j}$ which is depicted 
in $(b)$. The values in the right--hand column and in the bottom row are
the constraints $(A.13)$ which must be satisfied by the elements, respectively, 
of each row and each column of the matrix.}
}}

\vfill\eject
\bye